\documentstyle[12pt]{article}
\author{
Petr CHVOSTA\footnotemark[1]$\,\,$ and No\"{e}lle POTTIER\footnotemark[2]\\
Groupe de Physique des Solides\footnotemark[3],\\
Tour 23, 2 place Jussieu, 75251 Paris Cedex 05, France}
\title{{\bf One-Dimensional Diffusion in a Semi-Infinite Poisson Random Force}}
\date{November 16, 1998}
\addtocounter{footnote}{0}
\footnotetext{Permanent address: Department of Polymer Physics, 
Faculty of Mathematics and Physics, Charles University, 
V~Hole\v{s}ovi\v{c}k\'{a}ch 2, 180 00 Prague 8, Czech Republic.\\
Tel.: +4202-2191 2356; Fax: +4202-688 5095; 
e-mail: chvosta@kfpy.troja.mff.cuni.cz}
\addtocounter{footnote}{1}
\footnotetext{Tel.: +33-1-44274609; Fax: +33-1-43542878; 
e-mail: pottier@gps.jussieu.fr}
\addtocounter{footnote}{1}
\footnotetext{Laboratoire associ\'{e} au C.N.R.S. (UMR 7588) et aux 
Universit\'{e}s Paris VI et Paris VII.}
\addtocounter{footnote}{1}
\newcommand{\ttilde}{\mbox{$\widetilde{t}$}}
\newcommand{\xtilde}{\mbox{$\widetilde{x}$}}
\newcommand{\ytilde}{\mbox{$\widetilde{y}$}}

\newcommand{\Ptilde}{\mbox{$\widetilde{P}$}}
\newcommand{\Jtilde}{\mbox{$\widetilde{J}$}}

\newcommand{\sow}{\mbox{${\sf W}$}}
\newcommand{\soh}{\mbox{${\sf H}$}}
\newcommand{\soi}{\mbox{${\sf I}$}}
\newcommand{\sob}{\mbox{${\cal B}$}}
\newcommand{\sof}{\mbox{${\cal F}$}}

\newcommand{\sok}{\mbox{${\cal K}$}}

\begin{document}
\maketitle
\baselineskip 3.3ex
\footskip 5ex
\parindent 2.5em
\abovedisplayskip 5ex
\belowdisplayskip 5ex
\abovedisplayshortskip 3ex
\belowdisplayshortskip 5ex
\textfloatsep 7ex
\intextsep 7ex
%%%%%%%%%%%%%%%%%%%%%%%%%%%%%%%%%%%%%%%%%%%%%%%%%%%%%%%%%%%%%%%%%%%%%
%%%%%%%%%%%%%%%%%%%%%%%%%%%%%%%%%%%%%%%%%%%%%%%%%%%%%%%%%%%%%%%%%%%%%
%%%%%%%%%%%%%%%%%%%%%%%%%%%%%%%%%%%%%%%%%%%%%%%%%%%%%%%%%%%%%%%%%%%%%
\begin{abstract}
We consider the one-dimensional diffusion of a particle on a semi-infinite 
line and in a piecewise linear random potential. We first present a new 
formalism which yields an analytical expression for the Green function of the 
Fokker-Planck equation, valid for any deterministic construction of the 
potential profile. The force is then taken to be an asymmetric dichotomic 
process. Solving the corresponding energy-dependent stochastic Riccati 
equation in the space-asymptotic regime, we give an exact probabilistic 
description of returns to the origin. This method allows for a time-asymptotic 
characterization of the underlying dynamical phases. When the two values taken 
by the dichotomic force are of different signs, there occur trapping potential 
wells with a broad distribution of trapping times and dynamical phases may 
appear, depending on the mean force. If both values are negative, the 
time-asymptotic mean value of the probability density at the origin is 
proportional to the absolute value of the mean force. If they are both 
positive, traps no more exist and the dynamics is always normal. Problems with 
a shot-noise force and with a Gaussian white-noise force are solved as 
appropriate limiting cases.
\end{abstract}
%%%%%%%%%%%%%%%%%%%%%%%%%%%%%%%%%%%%%%%%%%%%%%%%%%%%%%%%%%%%%%%%%%%%%
\begin{center}
{\em Heading:\/} Statistical physics\\
{\em Short title:\/} Diffusion in Poisson Quenched Disorder 
Brownian motion.
\end{center}
\newpage
%%%%%%%%%%%%%%%%%%%%%%%%%%%%%%%%%%%%%%%%%%%%%%%%%%%%%%%%%%%%%%%%%%%%%%%
%%%%%%%%%%%%%%%%%%%%%%%%%%%%%%%%%%%%%%%%%%%%%%%%%%%%%%%%%%%%%%%%%%%%%%%
%%%%%%%%%%%%%%%%%%%%%%%%%%%%%%%%%%%%%%%%%%%%%%%%%%%%%%%%%%%%%%%%%%%%%%%
\section{Introduction}
%%%%%%%%%%%%%%%%%%%%%%%%%%%%%%%%%%%%%%%%%%%%%%%%%%%%%%%%%%%%%%%%%%%%%%%
%%%%%%%%%%%%%%%%%%%%%%%%%%%%%%%%%%%%%%%%%%%%%%%%%%%%%%%%%%%%%%%%%%%%%%%
%%%%%%%%%%%%%%%%%%%%%%%%%%%%%%%%%%%%%%%%%%%%%%%%%%%%%%%%%%%%%%%%%%%%%%%
The study of dynamical features of diffusion or conductivity in random 
environments has been initiated in early eighties \cite{alexander} and since 
then intensively pursued \cite{havlin}-\cite{bouchaud}. The problem is usually 
formulated either in a discrete form, {\em i.e.\/} by means of the Pauli 
Master Equation \cite{vankampen}, assuming the transfer rates to be random 
variables, or in a continuous formulation, {\em i.e.\/} using the 
Fokker-Planck equation \cite{gardiner}-\cite{horsthemke} and assuming the 
so-called drift-function to be a stochastic function in space. The transport 
properties are obtained through an average over the random parameters in the 
equation of motion. 

The resulting dynamics in the time-asymptotic region may exhibit non-standard 
features. For instance, with a Gaussian drift function of non-zero mean value, 
a succession of dynamical phases is observed when the bias is varied. These 
phases are characterized by different drift and diffusion behaviours, normal 
({\em i.e.\/} with a finite mobility and diffusion coefficient), or not. The 
existence of anomalous dynamical phases can be traced back to the existence of 
traps with a broad distribution of trapping times \cite{bouchaud}.

Most of the standard treatments assume independent random transfer rates in 
the master equation \cite{bernasconi1,bernasconi2}, or, correspondingly, a 
white-noise drift function in the Fokker-Planck equation 
\cite{vinokur}-\cite{aslangulg3}. Nevertheless, as pointed out {\em e.g.\/} in 
\cite{derrida}, a more realistic description should take into account the 
possibility of spatial correlations in the local transport parameters
\cite{aslangulcorr1}-\cite{comtet}. 

Let us for instance consider a continuous medium and imagine first that the 
particle diffuses through an array of randomly positioned minima with randomly 
distributed depths. The simplest choice for a space-correlated bias 
yielding this scenario is that of a dichotomic noise \cite{horsthemke}, 
with realizations alternately assuming two possible values of different signs. 
The potential then displays a succession of linear segments of random lengths 
and of alternately positive and negative slopes. One can well figure out that, 
in such a potential, the minima represent traps, some of them being deep traps 
with large trapping times. This case shares some similarities with the 
Gaussian-white-noise one, and should lead to results of the same type for the 
particle drift and diffusion properties ({\em i.e.\/} the existence of 
anomalous dynamical phases for certain values of the parameters of the model). 
But, interestingly enough, the choice of a dichotomic noise also allows to 
treat the case of a quenched random force taking alternately two values of the 
same sign. Since clearly the notion of deep traps---and even simply of traps---
makes no more sense in this case, one does not expect the existence of 
anomalous dynamical phases. 

More specifically, in the present paper, we assume that the lengths of the 
above described constant-force segments are independent and exponentially 
distributed, in which case the quenched random force is a Markovian Poisson 
process in space. We consider a particle diffusing on a semi-infinite line, 
{\em i.e.\/} we impose a reflecting boundary condition at the origin. These 
two hypotheses allow for exact analytical calculation and for the discussion 
of a rich variety of physically different situations. 

The paper is organized as follows. In Section 2, our analysis begins with the 
Laplace transformation of the Fokker-Planck equation. One thus gets a 
differential equation in space, depending parametrically on the energy, as 
pictured by the Laplace variable $z$. The corresponding Green function in the 
presence of an arbitrary deterministic piecewise constant bias is derived. In 
Section 3, we introduce quenched disorder with general piecewise constant 
realizations. The localization probability of the particle at its (sharp) 
initial position satisfies a Riccati stochastic differential equation with a 
multiplicative noise (in space). Specifying further the noise to be the 
Markovian Poisson one, we are then able to derive a one-formula based 
(Eq.\ (\ref{ran24})) parallel analysis of the disorder average of the 
localization probability of the particle at the origin on the one hand and the 
trapping time or the time-asymptotic average velocity on the other hand. In 
Section 4, we analyze various physical situations, according to the sign of 
the mean bias and to the presence or to the absence of traps. Finally, Section 
5 contains our conclusions. 

Generally speaking, the new results of our paper are the following. First, our 
procedure is exact for any fixed value of the Laplace variable $z$. On the one 
hand, this enables, at least in principle, the analysis of the 
disorder-averaged probability density at the origin for {\em any\/} time. On 
the other hand, we can carry out the small-$z$ analysis and derive the exact 
time-asymptotic formulae for this quantity in various physical regimes. 
Another specific feature of our work is the application of reflecting boundary 
conditions at the origin. Thus, clearly, in contrast to the equivalent problem 
on an infinite line, the situation with a positive or a negative mean bias 
{\em are not\/} equivalent. Indeed, with a negative mean bias the particle is 
in some sense stuck to the boundary at the origin or pushed back towards it, 
while with a positive mean bias it escapes towards infinity, the {\em modus\/}
of its time-asymptotic motion being controlled by a parameter describing 
the typical depth of the potential traps.
%%%%%%%%%%%%%%%%%%%%%%%%%%%%%%%%%%%%%%%%%%%%%%%%%%%%%%%%%%%%%%%%%%%%%%%
%%%%%%%%%%%%%%%%%%%%%%%%%%%%%%%%%%%%%%%%%%%%%%%%%%%%%%%%%%%%%%%%%%%%%%%
%%%%%%%%%%%%%%%%%%%%%%%%%%%%%%%%%%%%%%%%%%%%%%%%%%%%%%%%%%%%%%%%%%%%%%%
\section{Diffusion in a deterministic force}
%%%%%%%%%%%%%%%%%%%%%%%%%%%%%%%%%%%%%%%%%%%%%%%%%%%%%%%%%%%%%%%%%%%%%%%
%%%%%%%%%%%%%%%%%%%%%%%%%%%%%%%%%%%%%%%%%%%%%%%%%%%%%%%%%%%%%%%%%%%%%%%
%%%%%%%%%%%%%%%%%%%%%%%%%%%%%%%%%%%%%%%%%%%%%%%%%%%%%%%%%%%%%%%%%%%%%%%
\subsection{Homogeneous force}
%%%%%%%%%%%%%%%%%%%%%%%%%%%%%%%%%%%%%%%%%%%%%%%%%%%%%%%%%%%%%%%%%%%%%%%
%%%%%%%%%%%%%%%%%%%%%%%%%%%%%%%%%%%%%%%%%%%%%%%%%%%%%%%%%%%%%%%%%%%%%%%
Let us consider an overdamped Brownian particle acted upon by a standard 
white-noise Langevin force $\Gamma(\ttilde)$ and by a position-dependent 
potential force $F(\xtilde)$. Its dynamics is described by the viscous 
Langevin equation 
\begin{equation}
\label{det1}
\eta\frac{d}{d\ttilde}\,\xtilde(\ttilde)=
F[\xtilde(\ttilde)]+\Gamma(\ttilde)\,\,,
\end{equation}
with $\eta$ being the viscosity. The correlation function of the Langevin 
force is equal to $2D_{0}\eta^{2}\delta(\ttilde-\ttilde')$, where 
$D_{0}=k_{B}T/\eta$ is the diffusion constant in the absence of the potential 
force. The corresponding Fokker-Planck equation for the Green function 
$\Ptilde(\xtilde,\ytilde;\ttilde)$ reads
\begin{equation}
\label{det2}
\frac{\partial}{\partial\ttilde}\,\Ptilde(\xtilde,\ytilde;\ttilde)=
-\frac{\partial}{\partial\xtilde}
\left[\,
-D_{0}\frac{\partial}{\partial\xtilde}\,\Ptilde(\xtilde,\ytilde;\ttilde)+
\frac{F(\xtilde)}{\eta}\Ptilde(\xtilde,\ytilde;\ttilde)\,\right]\,\,.
\end{equation}
The bracketed expression represents the probability current 
$\Jtilde(\xtilde,\ytilde;\ttilde)$. We assume the initial condition 
$\Ptilde(\xtilde,\ytilde;0)=\delta(\xtilde-\ytilde)$ and the boundary 
conditions $\Jtilde(\xtilde_{0},\ytilde;t)=0$, 
$\Jtilde(\xtilde_{1},\ytilde;\ttilde)=0$. Consequently, the boundaries at 
$\xtilde_{0}$ and $\xtilde_{1}$ are reflecting and the probability density is 
always normalized to unity.

In order to make the following calculation more transparent, we introduce 
dimensionless variables. The potential force will be written in the form 
$F(\xtilde)=F_{0}f(\xtilde)$. The dimensionless coordinate is 
$x=\xtilde F_{0}/\eta D_{0}$, and the dimensionless time 
$t=\ttilde F_{0}^{2}/\eta^{2}D_{0}$. We thus get from Eq.\ (\ref{det2}):
\begin{eqnarray}
\label{det3}
\rule[-1ex]{0em}{4ex}\frac{\partial}{\partial t}P(x,y;t)&=&
-\frac{\partial}{\partial x}J(x,y;t)\,\,,\\
\label{det4}
\rule[-1ex]{0em}{4ex}J(x,y;t)&=&
-\frac{\partial}{\partial x}P(x,y;t)+f(x)P(x,y;t)\,\,.
\end{eqnarray}
The original density and current are connected with their dimensionless
counterparts {\em via\/} 
$\Ptilde(\xtilde,\ytilde;\ttilde)=F_{0}P(x,y;t)/\eta D$ and 
$\Jtilde(\xtilde,\ytilde;\ttilde)=F_{0}^{2}J(x,y;t)/\eta^{2}D$.

Let us assume for the moment a position-independent force: $f(x)=f$ for 
$x\in[x_{0},x_{1}]$. Performing the Laplace transformation, the Fokker-Planck 
equation (\ref{det2}) yields a nonhomogeneous differential equation with 
constant coefficients,
\begin{equation}
\label{det5}
\left[\frac{d^{2}}{dx^{2}}-f\frac{d}{dx}+z\right]\,P(x,y;z)=-\delta(x-y)\,\,.
\end{equation}
We are using the same symbol for a given function $a(t)$ and for its Laplace 
transform $a(z)=\int_{0}^{\infty}dt\,\exp(-zt) a(t)$, the Laplace original 
(transform) being always indicated by writing the variable $t$ ($z$). 
Combining any particular solution of the nonhomogeneous equation with the 
general solution of the homogeneous equation, we have 
$P(x,y;z)=P_{N}(x,y;z)+P_{H}(x;z)$, with 
\begin{eqnarray}
\label{det6}
\rule[-1ex]{0em}{4ex}P_{N}(x,y;z)&=&
\frac{1}{2\alpha(z)}\left\{\Theta(y-x){\rm e}^{(x-y)\alpha^{+}(z)}+
\Theta(x-y){\rm e}^{-(x-y)\alpha^{-}(z)}\right\}\,\,,\\
\label{det7}
\rule[-1ex]{0em}{4ex}P_{H}(x;z)&=&
c^{+}(z){\rm e}^{x\alpha^{+}(z)}+c^{-}(z){\rm e}^{-x\alpha^{-}(z)}\,\,,
\end{eqnarray}
where $\alpha(z)=\sqrt{z+f^{2}/4}$, $\alpha^{\pm}(z)=\sqrt{z+f^{2}/4}\pm f/2$, 
and $\Theta(x)$ is the Heaviside function. Having acquired the general 
solution for the density, we calculate the general expression for the 
probability current. In the last step, the two functions $c^{\pm}(z)$ are 
fixed from the reflecting-boundary conditions $J(x_{0},y;z)=0$, 
$J(x_{1},y;z)=0$.

The whole procedure is well known. However, in view of the following 
calculation, it is convenient to present the final result in a matrix form. We 
introduce a two-dimensional space with the basis 
$\{|1,0\rangle\,,\,|0,1\rangle\}$ and we express the pair density--current as 
two coordinates of a single state ket: $P(x,y;z)=\langle1,0|G(x,y;z)\rangle$ 
and $J(x,y;z)=\langle0,1|G(x,y;z)\rangle$. Adopting this convention, the 
result of the present simple example reads 
\begin{equation}
\label{det8}
|G(x,y;z)\rangle=
\sow(x_{1}-x;z)|1,0\rangle\,\Gamma(y;z)-\Theta(y-x)\sow(y-x;z)|0,1\rangle\,\,.
\end{equation}
Here we have introduced the abbreviation 
\begin{equation}
\label{det9}
\Gamma(y;z)=\frac{\langle0,1|\sow(y-x_{0};z)|0,1\rangle}
{\langle0,1|\sow(x_{1}-x_{0};z)|1,0\rangle}\,\,,
\end{equation}
and the matrix 
\begin{equation}
\label{det10}
\sow(x;z)=\left(\begin{array}{cc}
\rule[-2ex]{0em}{5ex}
\frac{\displaystyle
\alpha^{-}(z){\rm e}^{x\alpha^{-}(z)}+\alpha^{+}(z){\rm e}^{-x\alpha^{+}(z)}}
{\displaystyle2\alpha(z)}&
\frac{\displaystyle
{\rm e}^{x\alpha^{-}(z)}-{\rm e}^{-x\alpha^{+}(z)}}
{\displaystyle2\alpha(z)}\\
\rule[-2ex]{0em}{5ex}
z\,\frac{\displaystyle
{\rm e}^{x\alpha^{-}(z)}-{\rm e}^{-x\alpha^{+}(z)}}
{\displaystyle2\alpha(z)}&
\frac{\displaystyle
\alpha^{+}(z){\rm e}^{x\alpha^{-}(z)}+\alpha^{-}(z){\rm e}^{-x\alpha^{+}(z)}}
{\displaystyle2\alpha(z)}
\end{array}\right).
\end{equation}
Notice that, at given $z$, $\sow(x;z)$ satisfies the ``dynamical equation'' 
\begin{equation}
\label{det11}
\frac{d}{dx}\sow(x;z)=\soh(z)\sow(x;z)\hspace{2em},\hspace{2em}
\soh(z)=\left(\begin{array}{cc}-f&1\\z&0\end{array}\right)\,\,,
\end{equation}
with the position $x$ playing here the role of time. Eq.\ (\ref{det10}) 
gives $\sow(0;z)=\soi$ (unity matrix), {\em i.e.\/} we have formally 
$\sow(x;z)=\exp[x\soh(z)]$. 

Eq.\ (\ref{det8}) yields the complete picture of the resulting motion 
({\em i.e.\/} we can compute $P(x,y;z)$, $J(x,y;z)$ and these functions can be 
inverted into the time domain \cite{abramowitz}). For instance, taking 
$x_{0}=0$, $x_{1}=l$, $y\rightarrow 0^{+}$, and $x=0$, the probability density 
at the origin emerges as the ratio of two matrix elements: 
\begin{equation}
\label{det12}
P(0,0;z)=
\frac{\langle1,0|\sow(l;z)|1,0\rangle}{\langle0,1|\sow(l;z)|1,0\rangle}=
\frac{1}{z}\,\frac{\displaystyle
\alpha^{-}(z){\rm e}^{l\alpha^{-}(z)}+\alpha^{+}(z){\rm e}^{-l\alpha^{+}(z)}}
{\displaystyle{\rm e}^{l\alpha^{-}(z)}-{\rm e}^{-l\alpha^{+}(z)}}\,\,.
\end{equation}
Moreover, for the semi-infinite line, we have 
$\lim_{l\rightarrow\infty}P(0,0;z)=\alpha^{-}(z)/z$. In this case, we get 
following picture. Having $f<0$, the force pushes the diffusing particle 
against the boundary. In this case, the time-asymptotic value of the 
probability density at the origin is $|f|$, the asymptotic value of the mean 
particle position is $|f|^{-1}$, {\em i.e.\/} the time-asymptotic velocity is 
zero. On the other hand, when $f>0$, $\lim_{l\rightarrow\infty}P(0,0;t)$ 
decreases exponentially to zero and the mean position increases linearly with 
time, the velocity being just $f$. Finally, in the marginal case $f=0$, 
$P(0,0;t)$ behaves asymptotically as $1/\sqrt{\pi t}$, the mean position 
increases as $\sqrt{\pi t}$, and the asymptotic velocity is zero. 
%%%%%%%%%%%%%%%%%%%%%%%%%%%%%%%%%%%%%%%%%%%%%%%%%%%%%%%%%%%%%%%%%%%%%%%
%%%%%%%%%%%%%%%%%%%%%%%%%%%%%%%%%%%%%%%%%%%%%%%%%%%%%%%%%%%%%%%%%%%%%%%
\subsection{Piecewise constant force}
%%%%%%%%%%%%%%%%%%%%%%%%%%%%%%%%%%%%%%%%%%%%%%%%%%%%%%%%%%%%%%%%%%%%%%%
%%%%%%%%%%%%%%%%%%%%%%%%%%%%%%%%%%%%%%%%%%%%%%%%%%%%%%%%%%%%%%%%%%%%%%%
Let the original interval $[x_{0},x_{N}]$ be divided into $N$ segments 
$[x_{k-1},x_{k}]$, $k=1,\ldots,N$, with lengths $l_{k}=x_{k}-x_{k-1}$. Let 
$f_{k}$ be the constant force in the $k$-th subinterval. We assume that the 
particle has been originally placed in the $M$-th segment, {\em i.e.\/} 
$P(x,y;0)=\delta(x-y)$ with $y\in[x_{M-1},x_{M}]$. The boundary conditions at 
$x_{0}$ and $x_{N}$ are again reflecting. 

The procedure for solving the Fokker-Planck equation will be parallel to that 
in the above simple example. The general solution in the $M$-th segment 
consists of two parts. First, the particular solution of the nonhomogeneous 
equation (\ref{det5}) (with $f_{M}$ instead of $f$) will assume the form 
(\ref{det6}) with 
$\alpha_{M}(z)=\sqrt{z+f_{M}^{2}/4}$ instead of $\alpha(z)$ and 
$\alpha_{M}^{\pm}(z)=\alpha_{M}(z)\pm f_{M}/2$ instead of $\alpha^{\pm}(z)$. 
Second, the general solution of the homogeneous equation in the $M$-th 
subinterval assumes the form (\ref{det7}), again with 
$\alpha_{M}^{\pm}(z)$ instead of $\alpha^{\pm}(z)$, and with two arbitrary 
functions $c_{M}^{\pm}(z)$ instead of $c^{\pm}(z)$. The general solution in 
the $k$-th subinterval, $k\ne M$, is also of the form (\ref{det7}) with 
the substitutions $\alpha^{\pm}(z)\rightarrow\alpha_{k}^{\pm}(z)$ and 
$c^{\pm}(z)\rightarrow c_{k}^{\pm}(z)$. Altogether, the whole general solution 
depends on the $2N$ functions $c_{k}^{\pm}(z)$, $k=1,\ldots,N$. These are 
fixed from the requirements $J(x_{0},y;z)=J(x_{N},y;z)=0$ at the reflecting 
boundaries and from the continuity conditions for the probability density and 
for the probability current at the intermediate points 
$x_{1},x_{2},\ldots,x_{N-1}$. 

The final result can be again expressed in matrix form. We designate the 
``evolution operator'' for the $k$-th segment as $\sow_{k}(x;z)$---it is 
defined by the expression (\ref{det10}) with the substitutions 
$\alpha(z)\rightarrow\alpha_{k}(z)$ and 
$\alpha^{\pm}(z)\rightarrow\alpha_{k}^{\pm}(z)$. Further, we introduce the 
notations
\begin{eqnarray}
\label{det13}
\rule[-1ex]{0em}{4ex}\sow_{m,n}&=&
\sow_{m}(l_{m};z)\sow_{m+1}(l_{m+1};z)\ldots\sow_{n}(l_{n};z)\,\,,\\
\label{det14}
\rule[-1ex]{0em}{4ex}\Gamma_{M,N}(y;z)&=&
\frac{\langle0,1|\sow_{1,M-1}\sow_{M}(y-x_{M-1};z)|0,1\rangle}
{\langle0,1|\sow_{1,N}|1,0\rangle}\,\,.
\end{eqnarray}
Finally, let $|G_{k}(x,y;z)\rangle$ be the value of the state ket in the 
$k$-th segment, that is $|G(x,y;z)\rangle=|G_{k}(x,y;z)\rangle$ for 
$x\in[x_{k-1},x_{k}]$. The final result of the present Section then reads
\begin{eqnarray}
\rule[-1ex]{0em}{4ex}|G_{1}(x,y;z)\rangle&=&
\sow_{1}(x_{1}-x;z)\sow_{2,N}|1,0\rangle\,\Gamma_{M,N}(y;z)-\nonumber\\
\label{det15}
\rule[-1ex]{0em}{4ex}&-&
\sow_{1}(x_{1}-x;z)\sow_{2,M-1}\sow_{M}(y-x_{M-1};z)|0,1\rangle\,\,,\\
\rule[-1ex]{0em}{4ex}&\ldots&\nonumber\\
\rule[-1ex]{0em}{4ex}|G_{M-1}(x,y;z)\rangle&=&
\sow_{M-1}(x_{M-1}-x;z)\sow_{M,N}|1,0\rangle\,\Gamma_{M,N}(y;z)-\nonumber\\
\label{det16}
\rule[-1ex]{0em}{4ex}&-&
\sow_{M-1}(x_{M-1}-x;z)\sow_{M}(y-x_{M-1};z)|0,1\rangle\,\,,\\
\rule[-1ex]{0em}{4ex}|G_{M}(x,y;z)\rangle&=&
\sow_{M}(x_{M}-x;z)\sow_{M+1,N}|1,0\rangle\,\Gamma_{M,N}(y;z)-\nonumber\\
\label{det17}
\rule[-1ex]{0em}{4ex}&-&
\Theta(y-x)\sow_{M}(y-x;z)|0,1\rangle\,\,,\\
\label{det18}
\rule[-1ex]{0em}{4ex}|G_{M+1}(x,y;z)\rangle&=&
\sow_{M+1}(x_{M+1}-x;z)\sow_{M+2,N}|1,0\rangle\,\Gamma_{M,N}(y;z)\,\,,\\
\rule[-1ex]{0em}{4ex}&\ldots&\nonumber\\
\label{det19}
\rule[-1ex]{0em}{4ex}|G_{N}(x,y;z)\rangle&=&
\sow_{N}(x_{N}-x;z)|1,0\rangle\,\Gamma_{M,N}(y;z)\,\,.
\end{eqnarray}
It is easy to check that the boundary conditions and the continuity conditions 
are actually satisfied and that the resulting probability density is properly 
normalized. 

Having at hand the complete Green function for the given composition of the 
segments, we now proceed to the analysis of some consequences. First, let 
$x_{0}=0$, $x_{N}=l$, $y\rightarrow 0^{+}$, and $x=0$. The probability density 
at the origin assumes a particularly simple form:
\begin{equation}
\label{det20}
P(0,0;z)=
\frac{\langle1,0|\sow_{1}(l_{1};z)\sow_{2}(l_{2};z)\ldots\sow_{N}(l_{N};z)
|1,0\rangle}
{\langle0,1|\sow_{1}(l_{1};z)\sow_{2}(l_{2};z)\ldots\sow_{N}(l_{N};z)
|1,0\rangle}\,\,.
\end{equation}
In Section 3, this expression will be taken as a starting point for the 
disordered-medium calculation. 

Second, consider the value of the probability density at point $y$ in the 
$M$-th segment, {\em i.e.\/} at the initial position of the particle. Taking 
$x_{0}=0$ and $x\rightarrow y^{+}$, $P(x,x;z)$ equals the ratio
\begin{equation}
\label{det21}
\frac{\langle0,1|\sow_{1,M-1}\sow_{M}(x-x_{M-1};z)|0,1\rangle\,\,
\langle1,0|\sow_{M}(x_{M}-x;z)\sow_{M+1,N}|1,0\rangle}
{\langle0,1|\sow_{1}(l_{1};z)\ldots\sow_{M}(l_{M};z)\ldots\sow_{N}(l_{N};z)
|1,0\rangle}\,\,.
\end{equation}
Due to the exponential nature of the operator $\sow_{M}(l_{M};z)$ in the 
denominator, we can split it as a product of two factors, 
$\sow_{M}(x-x_{M-1};z)\sow_{M}(x_{M}-x;z)$. Further, we can insert in between 
the resolution of the unity operator, which yields 
\begin{eqnarray}
\rule[-1ex]{0em}{4ex}P(x,x;z)&=&\left[\frac{\langle0,1|
\sow_{k}(x_{k}-x;z)\sow_{k+1}(l_{k+1};z)\ldots\sow_{N}(l_{N};z)|1,0\rangle}
{\langle1,0|
\sow_{k}(x_{k}-x;z)\sow_{k+1}(l_{k+1};z)\ldots\sow_{N}(l_{N};z)|1,0\rangle}
+\right.\nonumber\\
\label{det22}
\rule[-1ex]{0em}{4ex}&+&\left.\frac{\langle0,1|
\sow_{1}(l_{1};z)\ldots\sow_{k-1}(l_{k-1};z)\sow_{k}(x-x_{k-1};z)|1,0\rangle}
{\langle0,1|
\sow_{1}(l_{1};z)\ldots\sow_{k-1}(l_{k-1};z)\sow_{k}(x-x_{k-1};z)|0,1\rangle}
\right]^{-1}.
\end{eqnarray}
For any {\em finite\/} $x$, the small-$z$ limit of the second term turns out 
to be zero. Thus the quantity $\lim_{z\rightarrow0^{+}}P(x,x;z)$ is equal to 
the small-$z$ limit of the density at origin for a {\em new\/} medium of the 
total length $l-x$. The new interval consists of $N-k+1$ segments of the 
lengths $x_{k}-x,l_{k+1},\ldots,l_{N}$, the constant forces in these segments 
being $f_{k},\ldots,f_{N}$. This simple result will be of considerable value 
in the analysis of the disordered medium. Namely, imagine that the particle is 
launched from the point $x$. The total time $T(x)$ spent in the interval 
$(x,x+dx)$ is equal to 
\begin{equation}
\label{det23}
T(x)\,dx=\left[\int_{0}^{\infty}P(x,x;t)\,dt\right]\,dx=
\lim_{z\rightarrow 0^{+}}P(x,x;z)\,dx\,\,,
\end{equation}
{\em i.e.\/} $T(x)$ can be related to the probability density at the origin 
for the above mentioned new interval. Of course, $T(x)$ can only be finite for 
an interval of infinite length. However, even for the semi-infinite interval, 
$T(x)$ is finite only if the particle can escape to infinity, that is, if the 
force in the last segment (which is necessarily of infinite length) is 
non-negative. 

Third, let us consider the thermally-averaged position of the particle 
$M(l;t)$ (we take again $x_{0}=0$, $x_{N}=l$, and $y\rightarrow 0^{+}$). 
Introducing the Laplace transform $M(l;z)=\int_{0}^{l}dx\,xP(x,0;z)$ and 
integrating the Laplace transform of the Fokker-Planck equation, we have 
\begin{eqnarray}
\rule[-1ex]{0em}{4ex}M(l;z)&=&\left[x\int_{0}^{x}\,dx'P(x',0;z)\right]_{0}^{l}-
\int_{0}^{l}dx\,\int_{0}^{x}dx'\,P(x',0;z)=\nonumber\\
\label{det24}
\rule[-1ex]{0em}{4ex}&&=\frac{l}{z}-\int_{0}^{l}dx\,\frac{1}{z}[1-J(x,0;z)]=
\frac{1}{z}\int_{0}^{l}\,J(x,0;z)\,dx\,\,.
\end{eqnarray}
Thereupon, taking the projection of the above Green function, we can write 
\begin{equation}
\label{det25}
M(l;z)=\frac{1}{z}\,
\frac{\displaystyle\sum_{k=1}^{N}\int_{x_{k-1}}^{x_{k}}\langle0,1|
\sow_{k}(x_{k}-x;z)\sow_{k+1}(l_{k+1};z)\ldots\sow_{N}(l_{N};z)|0,1\rangle\,dx}
{\displaystyle\langle0,1|
\sow_{1}(l_{1};z)\sow_{2}(l_{2};z)\ldots\sow_{N}(l_{N};z)|0,1\rangle}\,\,.
\end{equation}
In the next Section, this result will allow to connect the thermally-averaged 
position $M(l;z)$ with the Laplace transform $P(0,0;z)$ of the probability 
density at the origin. We now turn to the detailed analysis of this latter 
quantity in the disordered medium.
%%%%%%%%%%%%%%%%%%%%%%%%%%%%%%%%%%%%%%%%%%%%%%%%%%%%%%%%%%%%%%%%%%%%%%%
%%%%%%%%%%%%%%%%%%%%%%%%%%%%%%%%%%%%%%%%%%%%%%%%%%%%%%%%%%%%%%%%%%%%%%%
%%%%%%%%%%%%%%%%%%%%%%%%%%%%%%%%%%%%%%%%%%%%%%%%%%%%%%%%%%%%%%%%%%%%%%%
\section{Piecewise constant random force}
%%%%%%%%%%%%%%%%%%%%%%%%%%%%%%%%%%%%%%%%%%%%%%%%%%%%%%%%%%%%%%%%%%%%%%%
%%%%%%%%%%%%%%%%%%%%%%%%%%%%%%%%%%%%%%%%%%%%%%%%%%%%%%%%%%%%%%%%%%%%%%%
%%%%%%%%%%%%%%%%%%%%%%%%%%%%%%%%%%%%%%%%%%%%%%%%%%%%%%%%%%%%%%%%%%%%%%%
The preceding calculation in valid for any {\em deterministic\/} piecewise 
constant force. Such a force can be conceived as a member of a randomly 
constructed family of functions, that is, as a realization of a stochastic 
process. An arbitrary fixed realization is associated with a given weight. 
Consequently, the same weight is attributed to the probability density at the 
origin $P(0,0;z)$ for this realization. We are thus guided to the question: 
what is the probability density of the random variable $P(0,0;z)$? This 
Section presents the exact answer for a semi-infinite medium and a special 
type of the stochastic process to be fully specified in Subsection 3.3. In the 
first and the second Subsection, our reasoning is valid for {\em any\/} 
piecewise constant random force. 
%%%%%%%%%%%%%%%%%%%%%%%%%%%%%%%%%%%%%%%%%%%%%%%%%%%%%%%%%%%%%%%%%%%%%%%
%%%%%%%%%%%%%%%%%%%%%%%%%%%%%%%%%%%%%%%%%%%%%%%%%%%%%%%%%%%%%%%%%%%%%%%
\subsection{Stochastic Riccati equation}
%%%%%%%%%%%%%%%%%%%%%%%%%%%%%%%%%%%%%%%%%%%%%%%%%%%%%%%%%%%%%%%%%%%%%%%
%%%%%%%%%%%%%%%%%%%%%%%%%%%%%%%%%%%%%%%%%%%%%%%%%%%%%%%%%%%%%%%%%%%%%%%
Let us consider the following system of two stochastic differential equations,
\begin{equation}
\label{ran1}
\frac{d}{d\lambda}|\psi(\lambda;z)\rangle=
\soh(\lambda;z)|\psi(\lambda;z)\rangle\hspace{2em},\hspace{2em}
\soh(\lambda;z)=
\left(\begin{array}{cc}-\phi(\lambda)&1\\z&0\end{array}\right)\,\,,
\end{equation}
where $\phi(\lambda)$, $\lambda\ge0$, is a piecewise constant random process. 
Let us introduce the projections 
$R(\lambda;z)=\langle1,0|\psi(\lambda;z)\rangle$, 
$S(\lambda;z)=\langle0,1|\psi(\lambda;z)\rangle$, and take the initial 
conditions $R(0;z)=1$, $S(0;z)=0$. The system (\ref{ran1}) can be solved 
for any specific realization of $\phi(\lambda)$. Actually, consider the 
composition described at the beginning of Subsection 2.2. ({\em c.f.\/} also 
Fig.\ 1), and let the ``time'' $\lambda$ equal $l-x$. We have 
$\phi(\lambda;z)=f_{N}$ for $\lambda\in[0,l_{N}]$. If $\lambda$ increases, the 
evolution is controlled by the operator $\sow_{N}(\lambda;z)$. At the end of 
this interval, the state $\sow_{N}(l_{N};z)|1,0\rangle$ represents the initial 
condition for the semigroup evolution in the succeeding interval 
$\lambda\in[l_{N},l_{N-1}+l_{N}]$. This evolution is governed by the operator 
$\sow_{N-1}(\lambda;z)$. Repeating this reasoning, the solution at the end of 
the $N$-th interval, {\em i.e.\/} at the point $\lambda=l$, reads 
\begin{eqnarray}
\label{ran2}
\rule[-1ex]{0em}{4ex}R(l;z)&=&\langle1,0|
\sow_{1}(l_{1};z)\sow_{2}(l_{2};z)\ldots\sow_{N}(l_{N};z)|1,0\rangle\,\,,\\
\label{ran3}
\rule[-1ex]{0em}{4ex}S(l;z)&=&\langle0,1|
\sow_{1}(l_{1};z)\sow_{2}(l_{2};z)\ldots\sow_{N}(l_{N};z)|1,0\rangle\,\,.
\end{eqnarray}
By comparing with Eq.\ (\ref{det20}), it can be seen that the random variable 
$P(0,0;z)$ is identical to the ratio $R(l;z)/S(l;z)$. From now on, the random 
variable $P(0,0;z)$ will be designated as $P(\lambda;z)$. We introduce its 
probability density 
$\pi(p,\lambda;z)=\left\langle\,\delta[p-P(\lambda;z)]\,\right\rangle$, where 
the symbol $\langle\ldots\rangle$ denotes the average over the quenched 
disorder. Taking the projections of Eq.\ (\ref{ran1}), the random 
variables $R(\lambda;z)$ and $S(\lambda;z)$ obey the system of stochastic 
differential equations 
\begin{equation}
\label{ran4}
\frac{d}{d\lambda}R(\lambda;z)=-\phi(\lambda)R(\lambda;z)+S(\lambda;z)
\,\,\,\,,\,\,\,\,
\frac{d}{d\lambda}S(\lambda;z)=zR(\lambda;z)\,\,,
\end{equation}
with $\lambda$ representing the total length of the interval accessible to 
the diffusing particle. Finally, on comparing the $\lambda$-derivative of the 
product $R(\lambda;z)=P(\lambda;z)\,S(\lambda;z)$ with the first equation 
(\ref{ran4}) and on dividing by $S(\lambda;z)$, we obtain the stochastic 
Riccati differential equation obeyed by $P(\lambda;z)$:
\begin{equation}
\label{ran5}
\frac{d}{d\lambda}P(\lambda;z)=-zP^{2}(\lambda;z)-\phi(\lambda)P(\lambda;z)+1
\hspace{1em},\hspace{1em}P(0;z)=+\infty\,\,.
\end{equation}
For purely operational reasons, we introduce the random variable 
$Q(\lambda;z)=zP(\lambda;z)$. One gets from Eq.\ (\ref{ran5}):
\begin{equation}
\label{ran6}
\frac{d}{d\lambda}Q(\lambda;z)=-Q^{2}(\lambda;z)-\phi(\lambda)Q(\lambda;z)+z
\hspace{1em},\hspace{1em}Q(0;z)=+\infty\,\,.
\end{equation}
The density 
$\kappa(q,\lambda;z)=\left\langle\,\delta[q-Q(\lambda;z)]\,\right\rangle$ is 
related with the density $\pi(p,\lambda;z)$ by 
$\pi(p,\lambda;z)=z\kappa(zp,\lambda;z)$. Regarding the relation 
\begin{equation}
\label{ran7}
\lim_{t\rightarrow\infty}
P(0,0;t)=\lim_{z\rightarrow0^{+}}zP(\lambda;z)=
\lim_{z\rightarrow0^{+}}Q(\lambda;z)\,\,,
\end{equation}
the small-$z$ limit of the density $\kappa(q,\lambda;z)$ describes the 
time-asymptotic properties of the random variable $P(0,0;t)$.
%%%%%%%%%%%%%%%%%%%%%%%%%%%%%%%%%%%%%%%%%%%%%%%%%%%%%%%%%%%%%%%%%%%%%%%
%%%%%%%%%%%%%%%%%%%%%%%%%%%%%%%%%%%%%%%%%%%%%%%%%%%%%%%%%%%%%%%%%%%%%%%
\subsection{Mean trapping time and mean velocity}
%%%%%%%%%%%%%%%%%%%%%%%%%%%%%%%%%%%%%%%%%%%%%%%%%%%%%%%%%%%%%%%%%%%%%%%
%%%%%%%%%%%%%%%%%%%%%%%%%%%%%%%%%%%%%%%%%%%%%%%%%%%%%%%%%%%%%%%%%%%%%%%
Let us now return to the reasoning associated with Eqs.\ 
(\ref{det21})-(\ref{det23}). We shall call the variable 
$T(x)=\lim_{z\rightarrow0^{+}}P(x,x;z)$ the 
{\em trapping time\/}\footnote[4]{We shall use this term even if the potential 
is monotonous, although a more appropriate name in this case would be the {\em 
dwelling time\/}.}. Presently, the density $P(x,x;z)$ and hence also the 
trapping time are random 
variables. Performing the small-$z$ limit in Eq.\ (\ref{det22}), we have 
connected the trapping time with the particle-position probability density at 
the origin of an interval shorter than the original one. However, if the 
original interval is of {\em infinite\/} length, the same is true for the new 
one. Since the quenched random force is described by a {\em stationary\/} 
process, the probability density at the beginning of the new interval is 
stochastically equivalent with the density at the beginning of the original 
one. Summing up, one has:
\begin{equation}
\label{ran8}
T(x)=\lim_{z\rightarrow0^{+}}\lim_{\lambda\rightarrow\infty}P(x,x;z)=
\lim_{z\rightarrow0^{+}}\lim_{\lambda\rightarrow\infty}P(\lambda;z)\,\,.
\end{equation}
The trapping time is position-independent, {\em i.e.\/} $T(x)=T$, its mean 
value being 
\begin{equation}
\label{ran9}
\tau\stackrel{{\rm def}}{=}
\lim_{z\rightarrow0^{+}}\lim_{\lambda\rightarrow\infty}
\langle P(\lambda;z)\rangle=
\lim_{z\rightarrow0^{+}}\lim_{\lambda\rightarrow\infty}
\frac{1}{z}\langle Q(\lambda;z)\rangle\,\,.
\end{equation}

In the preceding Section, we have introduced the Laplace transform of the 
thermally-averaged position, $M(\lambda;z)$. Presently, it is again a random 
variable. Using the definition of $S(\lambda;z)$, the probability current in 
Eq.\ (\ref{det25}) can be rewritten as $J(x,0;z)=S(\lambda-x;z)/S(\lambda;z)$, 
which yields 
\begin{equation}
\label{ran10}
M(\lambda;z)=\frac{1}{z}\int_{0}^{\lambda}dx J(x,0;z)=
\frac{1}{z}
\frac{\displaystyle\int_{0}^{\lambda}dx\,S(x;z)}{S(\lambda;z)}\,\,,
\end{equation}
where we have used the stationarity of the quenched random force. The 
derivative of this equation yields
\begin{equation}
\label{ran11}
\frac{d}{d\lambda}M(\lambda;z)=\frac{1}{z}-zP(\lambda;z)M(\lambda;z)
\hspace{2em},\hspace{2em}M(0;z)=0\,\,.
\end{equation}
Finally, introducing the Laplace transform of the thermally-averaged velocity, 
$V(\lambda;z)=zM(\lambda;z)$, the {\em time\/}-asymptotic velocity for the 
semi-infinite line is given by 
\begin{equation}
\label{ran12}
\lim_{t\rightarrow\infty}\lim_{\lambda\rightarrow\infty}V(\lambda;t)=
\lim_{z\rightarrow0^{+}}\lim_{\lambda\rightarrow\infty}
z\int_{0}^{\lambda}d\lambda'
\exp\left[-z\int_{\lambda'}^{\lambda}d\lambda''P(\lambda'';z)\right]\,\,,
\end{equation}
where we have used 
$\lim_{t\rightarrow\infty}\lim_{\lambda\rightarrow\infty}V(\lambda;t)=
\lim_{z\rightarrow0^{+}}\lim_{\lambda\rightarrow\infty}zV(\lambda;z)$. The 
last formula can be rewritten in the form which reveals the well known 
\cite{bouchaud,aslangulg1} {\em self-averaging\/} property of the 
time-asymptotic velocity, when it is nonzero. Indeed, {\em assuming\/} that 
the limit $\lim_{z\rightarrow0^{+}}\lim_{\lambda\rightarrow\infty} 
\langle P(\lambda;z)\rangle$ is finite, we can write 
\begin{eqnarray}
\rule[-1ex]{0em}{6ex}
\lim_{t\rightarrow\infty}\lim_{\lambda\rightarrow\infty}V(\lambda;t)&=&
\lim_{z\rightarrow0^{+}}\lim_{\lambda\rightarrow\infty}
z\int_{0}^{\lambda}d\lambda'
\exp\left[-z(\lambda-\lambda')\langle P(\lambda;z)\rangle\right]
\times\nonumber\\
\label{ran13}
\rule[-1ex]{0em}{6ex}&\times&
\exp\left\{-z\int_{\lambda'}^{\lambda}d\lambda''
\left[P(\lambda'';z)-\langle P(\lambda;z)\rangle\right]\right\}\,\,.
\end{eqnarray}
Due to the above assumption, if $\lambda\rightarrow\infty$, the integral in 
the second exponent represents a random variable which is finite (for high 
enough $\lambda''$, the typical trajectory of $P(\lambda'';z)$ swings around 
the mean value $\langle P(\lambda;z)\rangle$). Thereupon, for small $z$, the 
second exponential tends to unity and the remaining integration yields the 
reciprocal value of the non-random number (\ref{ran9}). Thus, when the mean 
trapping time $\tau$ is finite, the asymptotic velocity is a self-averaging 
quantity equal to $\tau^{-1}$ \footnote[5]{For the exact proof of this 
statement one needs the simultaneous probability density $\psi(p,w,\lambda;z)$ 
for the random variables $P(\lambda;z)$ and $W(\lambda;z)=z^{2}M(\lambda;z)$. 
Assuming the condition in the text, one has to prove 
$\lim_{z\rightarrow0^{+}}\lim_{\lambda\rightarrow\infty} 
\int\psi(p,w,\lambda;z)\,dp=\delta(w-\tau^{-1})$.}. If the mean trapping 
time diverges, the asymptotic velocity vanishes. In this latter case, the 
disorder-averaged time-asymptotic mean position either tends to a constant 
(for a negative mean force), or increases slower than linearly. 

For the sake of completeness, the {\em second\/} thermally-averaged moment 
$N(\lambda;z)=\int_{0}^{\lambda}dx\,x^{2}P(x,0;z)$ can be also connected to 
the probability density at the origin. Actually, first, the Fokker-Planck 
equation implies $N(\lambda;z)=(2/z)\int_{0}^{\lambda}dx\,xJ(x,0;z)$. 
Thereupon, on deriving this expression, we get
\begin{equation}
\label{ran11a}
\frac{d}{d\lambda}N(\lambda;z)=2M(\lambda;z)-zP(\lambda;z)N(\lambda;z)
\hspace{2em},\hspace{2em}N(0;z)=0\,\,.
\end{equation}
Summing up, the first and the second thermally-averaged moments obey a system 
of stochastic differential equations (\ref{ran11}), (\ref{ran11a}), with 
$P(\lambda;z)$ playing the role of the ``input'' noise. 
%%%%%%%%%%%%%%%%%%%%%%%%%%%%%%%%%%%%%%%%%%%%%%%%%%%%%%%%%%%%%%%%%%%%%%%
%%%%%%%%%%%%%%%%%%%%%%%%%%%%%%%%%%%%%%%%%%%%%%%%%%%%%%%%%%%%%%%%%%%%%%%
\subsection{Dichotomic random force}
%%%%%%%%%%%%%%%%%%%%%%%%%%%%%%%%%%%%%%%%%%%%%%%%%%%%%%%%%%%%%%%%%%%%%%%
%%%%%%%%%%%%%%%%%%%%%%%%%%%%%%%%%%%%%%%%%%%%%%%%%%%%%%%%%%%%%%%%%%%%%%%
Let the forces in the individual segments assume alternately just two values, 
$F_{\pm}=F_{0}f_{\pm}$. We shall always take $f_{-}<f_{+}$; the equality would 
imply a position-independent constant force. Let the lengths of the 
constant-force segments be independent random variables. The generic 
probability density for the (dimensionless) lengths of the constant-force 
segments will be taken of the form 
$\rho_{\pm}(\lambda)=n_{\pm}\exp(-\lambda n_{\pm})$, where $1/n_{\pm}$ denotes 
the mean length of the segments with the force $f_{\pm}$. Due to this 
assumption, the resulting four-parameter stochastic process $\phi(\lambda)$ is 
Markovian and it usually referred to as the {\em asymmetric dichotomic 
noise\/} \cite{horsthemke}. We shall always work with a {\em stationary\/} 
dichotomic noise. We can set
\begin{equation}
\label{ran14}
\mu\stackrel{{\rm def}}{=}\langle\phi(\lambda)\rangle=
\frac{f_{-}n_{+}+f_{+}n_{-}}{n_{-}+n_{+}}\,\,\,\,,\,\,\,\,
\langle\,\phi(\lambda)\phi(\lambda')\,\rangle=
\frac{\sigma}{\lambda_{c}}\exp\left(-\frac{|\lambda-\lambda'|}{\lambda_{c}}
\right)\,\,, 
\end{equation}
where we have introduced the intensity 
$\sigma\stackrel{{\rm def}}{=}n_{-}n_{+}(f_{+}-f_{-})^{2}/(n_{-}+n_{+})^{3}$, 
and the correlation length
$\lambda_{c}\stackrel{{\rm def}}{=}(n_{-}+n_{+})^{-1}$. The statistical 
properties of the stationary noise are invariant with respect to the inversion 
$\lambda\rightarrow-\lambda$ and to the translation 
$\lambda\rightarrow\lambda-l$.

Let us now focus on Eq.\ (\ref{ran6}). It is convenient to associate 
with the variable $Q(\lambda;z)$ an overdamped motion of a hypothetical 
particle. While the ``time'' $\lambda$ increases, this particle moves 
alternately under the influence of the ``forces''
\begin{equation}
\label{ran15}
K_{\pm}(q;z)=-q^{2}-f_{\pm}q+z=-[q-q_{\pm}(z)]\,[q-q_{\pm}'(z)]\,\,,\\
\end{equation}
where we have introduced the four quantities 
\begin{equation}
\label{ran16}
q_{\pm}(z)=\sqrt{z+f_{\pm}^{2}/4}-f_{\pm}/2
\hspace{1em},\hspace{1em}
q_{\pm}'(z)=-\sqrt{z+f_{\pm}^{2}/4}-f_{\pm}/2\,\,.
\end{equation}
Notice the ordering $q_{+}'(z)<q_{-}'(z)<0<q_{+}(z)<q_{-}(z)$, valid for any 
real positive $z$ and for any values of the parameters $f_{-}<f_{+}$. The 
corresponding ``potentials'' $U_{\pm}(q;z)=-\int\,K_{\pm}(q;z)\,dq$ display 
minima at $q_{\pm}(z)$ and maxima at $q_{\pm}'(z)$. Starting from its initial 
``position'' at infinity, the particle always slides either towards 
$q_{-}(z)$ or towards $q_{+}(z)<q_{-}(z)$. For any fixed $\lambda$ the 
particle can only be found between the coordinate valid for the potential 
$U_{+}(q;z)$ and that valid for $U_{-}(q;z)$. Accordingly, for an arbitrary 
nonzero $z$, the probability density $\kappa(q,\lambda;z)$ vanishes outside 
the finite interval
\begin{equation}
\label{ran17}
\left[
\frac{\displaystyle
q_{+}(z){\rm e}^{\lambda q_{+}(z)}-q_{+}'(z){\rm e}^{\lambda q_{+}'(z)}}
{\displaystyle{\rm e}^{\lambda q_{+}(z)}-{\rm e}^{\lambda q_{+}'(z)}}
\,\,,\,\,
\frac{\displaystyle
q_{-}(z){\rm e}^{\lambda q_{-}(z)}-q_{-}'(z){\rm e}^{\lambda q_{-}'(z)}}
{\displaystyle{\rm e}^{\lambda q_{-}(z)}-{\rm e}^{\lambda q_{-}'(z)}}
\right]\,\,,
\end{equation}
and it displays two $\delta$-function contributions at the edges of this 
support, their weights being $n_{\mp}\exp(-\lambda n_{\pm})/(n_{-}+n_{+})$. 
This singular part describes an exponentially decreasing probability of having 
just one segment in the whole interval of length $\lambda$. Obviously, in the 
limit $\lambda\rightarrow\infty$, the support is simply $[q_{+}(z),q_{-}(z)]$ 
and the singular part is missing. 

In order to solve Eq.\ (\ref{ran6}) for the dichotomic noise in question
we follow the standard steps as described in \cite{horsthemke}. First, we 
introduce the joint densities 
\begin{equation}
\label{ran18}
\kappa_{\pm}(q,\lambda;z)\,dq=
{\rm Prob}\left\{\,Q(\lambda;z)\in(q,q+dq)\,\,{\rm and}\,\, \phi(\lambda)=
f_{\pm}\right\}\,\,. 
\end{equation}
One has $\kappa(q,\lambda;z)=\kappa_{-}(q,\lambda;z)+\kappa_{+}(q,\lambda;z)$. 
These densities obey the coupled partial differential equations
\begin{eqnarray}
\rule[-1ex]{0em}{6ex}\frac{\partial}{\partial\lambda}
\left[\begin{array}{c}
\kappa_{-}(q,\lambda;z)\\\kappa_{+}(q,\lambda;z)
\end{array}\right]&=&-
\frac{\partial}{\partial q}
\left[\begin{array}{c}
K_{-}(q;z)\kappa_{-}(q,\lambda;z)\\K_{+}(q;z)\kappa_{+}(q,\lambda;z)
\end{array}\right]-\nonumber\\
\label{ran19}
\rule[-1ex]{0em}{6ex}&-&
\left[\begin{array}{cc}n_{-}&-n_{+}\\-n_{-}&n_{+}\end{array}\right]\,
\left[\begin{array}{c}
\kappa_{-}(q,\lambda;z)\\\kappa_{+}(q,\lambda;z)
\end{array}\right]\,.
\end{eqnarray}
We are looking for the stationary solution 
$\kappa_{\pm}(q;z)=\lim_{\lambda\rightarrow\infty}\kappa_{\pm}(q,\lambda;z)
$\footnote[7]{The nonstationary case would describe the diffusion on a finite 
interval, its analysis being very difficult even in a simpler context 
\cite{shibani} and of minor physical importance.}. Hence we remove the 
$\lambda$-derivative on the l.h.s.\ of Eq.\ (\ref{ran19}). 
Introducing the two new functions
\begin{eqnarray}
\label{ran20}
\rule[-1ex]{0em}{6ex}\xi(q;z)&=&
\frac{K_{-}(q;z)K_{+}(q;z)}{n_{-}K_{+}(q;z)+n_{+}K_{-}(q;z)}
\left[n_{-}\kappa_{-}(q;z)-n_{+}\kappa_{+}(q;z)\right]\,\,,\\
\label{ran21}
\rule[-1ex]{0em}{6ex}\eta(q;z)&=&
K_{-}(q;z)\kappa_{-}(q;z)+K_{+}(q;z)\kappa_{+}(q;z)\,\,,
\end{eqnarray}
and carrying out the corresponding substitution in Eq.\ (\ref{ran19}), we
arrive at two independent equations: 
\begin{equation}
\label{ran22}
\frac{d}{dq}\eta(q;z)=0\hspace{1em},\hspace{1em}
\frac{1}{\xi(q;z)}\frac{d}{dq}\xi(q;z)=-
\left(\frac{n_{-}}{K_{-}(q;z)}+\frac{n_{+}}{K_{+}(q;z)}\right)\,\,.
\end{equation}
Hence the function $\eta(q;z)$ is simply a constant, which in fact is equal to 
zero. Actually, Eq.\ (\ref{ran15}) yields $K_{\pm}[q_{\pm}(z);z]=0$, and 
conservation of probability entails $\kappa_{\pm}[q_{\mp}(z);z]=0$. As for the 
differential equation for $\xi(q;z)$, its solution reads
\begin{equation}
\label{ran23}
\xi(q;z)=\frac{1}{C(z)}\left[\frac{q_{-}(z)-q}{q-q_{-}'(z)}\right]^{\nu_{-}(z)}
\left[\frac{q-q_{+}(z)}{q-q_{+}'(z)}\right]^{\nu_{+}(z)}
\hspace{1em},\hspace{1em}
\nu_{\pm}(z)=\frac{n_{\pm}}{\sqrt{4z+f_{\pm}^{2}}}\,\,,
\end{equation}
where $C(z)$ is a normalization constant. Finally, inverting the 
transformation $\{$(\ref{ran20}), (\ref{ran21})$\}$, we get 
$\kappa_{\pm}(q;z)=\mp\xi(q;z)/K_{\pm}(q;z)$ and the stationary density 
$\kappa(q;z)=\lim_{\lambda\rightarrow\infty}\kappa(q,\lambda;z)$ reads
\begin{eqnarray}
\rule[-1ex]{0em}{6ex}\kappa(q;z)&=&\frac{1}{C(z)}
\left\{
\frac{1}{[q_{-}(z)-q][q-q_{-}'(z)]}+\frac{1}{[q-q_{+}(z)][q-q_{+}'(z)]}
\right\}\times\nonumber\\
\label{ran24}
\rule[-1ex]{0em}{6ex}&\times&
\left[\frac{q_{-}(z)-q}{q-q_{-}'(z)}\right]^{\nu_{-}(z)}
\left[\frac{q-q_{+}(z)}{q-q_{+}'(z)}\right]^{\nu_{+}(z)}
\Theta[q;q_{+}(z),q_{-}(z)]\,\,,
\end{eqnarray}
where we have denoted\footnote[8]{The product of the two Heaviside functions 
guarantees the vanishing of the probability density outside the interval 
$[q_{+}(z),q_{-}(z)]$.} $\Theta(q;x,y)=\Theta(q-x)\Theta(y-q)$. In the final 
step, the constant $C(z)$ is fixed from the normalization condition 
$\int_{q_{+}(z)}^{q_{-}(z)}\kappa(q;z)\,dq=1$, that is 
\begin{equation}
\label{ran25}
C(z)=(f_{+}-f_{-})\displaystyle\int_{q_{+}(z)}^{q_{-}(z)}\,q\,
\frac{[q_{-}(z)-q]^{\nu_{-}(z)-1}\,[q-q_{+}(z)]^{\nu_{+}(z)-1}}
{[q-q_{-}'(z)]^{\nu_{-}(z)+1}\,[q-q_{+}'(z)]^{\nu_{+}(z)+1}}\,dq\,\,.
\end{equation}
Some steps of the underlying calculation are given in the Appendix---the 
resulting form of the above integral can be written as 
\begin{eqnarray}
\rule[-1ex]{0em}{6ex}C(z)&=&
\frac{1}{\left\langle\sqrt{4z+\phi^{2}(\lambda)}\right\rangle}\,\,
\frac{[q_{-}(z)-q_{+}(z)]^{\nu_{-}(z)+\nu_{+}(z)}}
{[q_{+}(z)-q_{-}'(z)]^{\nu_{-}(z)}[q_{-}(z)-q_{+}'(z)]^{\nu_{+}(z)}}
\times\nonumber\\
\label{ran26}
\rule[-1ex]{0em}{6ex}&\times&
\sob[\nu_{-}(z),\nu_{-}(z)]\,\,
\sof\left[\nu_{-}(z),\nu_{+}(z),\nu_{-}(z)+\nu_{+}(z)+1;-u(z)\right].
\end{eqnarray}
Here $\sob(x,y)=\Gamma(x)\Gamma(y)/\Gamma(x+y)$ denotes the Euler beta 
function, $\sof(a,b,c;x)$ is the Gauss hypergeometric function 
\cite{abramowitz}, and we have used the abbreviation
\begin{equation}
\label{ran27}
u(z)=\frac{[q_{-}(z)-q_{+}(z)][q_{+}'(z)-q_{+}'(z)]}
{[q_{-}(z)-q_{+}'(z)][q_{+}'(z)-q_{-}'(z)]}\,\,.
\end{equation}
Eqs.\ (\ref{ran24}), (\ref{ran26}) for the stationary density represent the 
main result of the present Section. The corresponding moments can be obtained 
by the usual integration. Again, the ensuing calculation is commented on in 
the Appendix.
%%%%%%%%%%%%%%%%%%%%%%%%%%%%%%%%%%%%%%%%%%%%%%%%%%%%%%%%%%%%%%%%%%%%%%%
%%%%%%%%%%%%%%%%%%%%%%%%%%%%%%%%%%%%%%%%%%%%%%%%%%%%%%%%%%%%%%%%%%%%%%%
%%%%%%%%%%%%%%%%%%%%%%%%%%%%%%%%%%%%%%%%%%%%%%%%%%%%%%%%%%%%%%%%%%%%%%%
\section{Discussion}
%%%%%%%%%%%%%%%%%%%%%%%%%%%%%%%%%%%%%%%%%%%%%%%%%%%%%%%%%%%%%%%%%%%%%%%
%%%%%%%%%%%%%%%%%%%%%%%%%%%%%%%%%%%%%%%%%%%%%%%%%%%%%%%%%%%%%%%%%%%%%%%
%%%%%%%%%%%%%%%%%%%%%%%%%%%%%%%%%%%%%%%%%%%%%%%%%%%%%%%%%%%%%%%%%%%%%%%
Our general four-parameter description of the dichotomic force provides a rich 
spectrum of special regimes, which can be analyzed using Eq.\ (\ref{ran24}), 
where, as indicated above, the probability density $\kappa(q;z)$ stands for 
$\lim_{\lambda\rightarrow\infty}\kappa(q,\lambda;z)$. In the same way, we 
shall use the simpler designations $Q(z)$, $\pi(p;z)$ and $P(z)$ for the 
stationary values $Q(\lambda;z)$, $\pi(p,\lambda;z)$ and $P(\lambda;z)$.
%%%%%%%%%%%%%%%%%%%%%%%%%%%%%%%%%%%%%%%%%%%%%%%%%%%%%%%%%%%%%%%%%%%%%%%
%%%%%%%%%%%%%%%%%%%%%%%%%%%%%%%%%%%%%%%%%%%%%%%%%%%%%%%%%%%%%%%%%%%%%%%
\subsection{Both forces are negative ($f_{-}<f_{+}<0$)}
%%%%%%%%%%%%%%%%%%%%%%%%%%%%%%%%%%%%%%%%%%%%%%%%%%%%%%%%%%%%%%%%%%%%%%%
%%%%%%%%%%%%%%%%%%%%%%%%%%%%%%%%%%%%%%%%%%%%%%%%%%%%%%%%%%%%%%%%%%%%%%%
The mean force $\mu$ in Eq.\ (\ref{ran14}) is negative and the potential 
consists of segments with a positive slope ({\em c.f.\/} Fig.\ 2). For any 
arbitrary realization of the quenched noise, the particle cannot escape to 
infinity---it can be found with probability one in a {\em finite\/} region. 
Intuitively, one expects a nonzero time-asymptotic mean value of the 
probability density at the origin, and a finite time-asymptotic value of the 
thermally-averaged mean position. 

Consider the small-$z$ limit of the probability density (\ref{ran24}). 
First, one has $\lim_{z\rightarrow 0^{+}}\nu_{\pm}(z)=n_{\pm}/|f_{\pm}|$. 
Second, the small-$z$ limits of the expressions $q_{\pm}(z)$ are $|f_{\pm}|$, 
whereas $q_{\pm}'(z)$ behave as $-z/|f_{\pm}|$. Analyzing the expressions in 
(\ref{ran17}), the support of the probability density 
$\lim_{z\rightarrow 0^{+}}\kappa(q;z)$ is the interval $[|f_{+}|,|f_{-}|]$. 
Finally, we have $u(z)\rightarrow 0$, {\em i.e.\/} the hypergeometric function 
in Eq.\ (\ref{ran26}) tends to unity. On collecting these observations, 
one gets
\begin{eqnarray}
\rule[-1ex]{0em}{6ex}\lim_{z\rightarrow0^{+}}\kappa(p;z)&=&|\mu|\,
\frac{|f_{-}|^{\frac{n_{+}}{|f_{+}|}}|f_{+}|^{\frac{n_{-}}{|f_{-}|}}}
{(|f_{-}|-|f_{+}|)^{\frac{n_{-}}{|f_{-}|}+\frac{n_{-}}{|f_{-}|}+1}}
\sob^{-1}\left(\frac{n_{-}}{|f_{-}|},\frac{n_{+}}{|f_{+}|}\right)
\times\nonumber\\
\label{dis1}
\rule[-1ex]{0em}{6ex}&\times&
\frac{(|f_{-}|-q)^{\frac{n_{-}}{|f_{-}|}-1}(q-|f_{+}|)^{\frac{n_{+}}{|f_{+}|}
-1}}{q^{\frac{n_{-}}{|f_{-}|}+\frac{n_{+}}{|f_{+}|}+1}}\,
\Theta(q;|f_{+}|,|f_{-}|)\,\,.
\end{eqnarray}
The corresponding moments 
$\lim_{z\rightarrow0^{+}}\left\langle Q^{k}(z)\right\rangle$ are all finite 
and they can be computed analytically by direct integration. In particular for 
$k=1$ one gets $\lim_{z\rightarrow0^{+}}\left\langle Q(z)\right\rangle=|\mu|$.
Thus the time-asymptotic mean value $\lim_{t\rightarrow\infty}\langle 
P(0,0;t)\rangle$ in the semi-infinite line is seen to be equal to the absolute 
value of the mean force, $|\mu|$. Note that this result cannot be obtained 
from the solution of the corresponding free-diffusion model on an infinite 
line ({\em i.e.\/} without the reflecting boundary condition at the origin).
The higher moments are not so simply related to the properties of the random 
force. Further, in the present case, the mean trapping time (\ref{ran9}) is 
infinite and the time-asymptotic velocity vanishes. The particle is in some 
sense stuck to the origin. If $f_{-}<f_{+}=0$, a slightly more complicated 
calculation reveals the same general conclusions.
%%%%%%%%%%%%%%%%%%%%%%%%%%%%%%%%%%%%%%%%%%%%%%%%%%%%%%%%%%%%%%%%%%%%%%%
%%%%%%%%%%%%%%%%%%%%%%%%%%%%%%%%%%%%%%%%%%%%%%%%%%%%%%%%%%%%%%%%%%%%%%%
\subsection{Forces are of different signs ($f_{-}<0<f_{+}$)}
%%%%%%%%%%%%%%%%%%%%%%%%%%%%%%%%%%%%%%%%%%%%%%%%%%%%%%%%%%%%%%%%%%%%%%%
%%%%%%%%%%%%%%%%%%%%%%%%%%%%%%%%%%%%%%%%%%%%%%%%%%%%%%%%%%%%%%%%%%%%%%%
The potential forms a system of traps ({\em c.f.\/} Fig.\ 1). The traps can 
only be efficient if the ratio $n_{+}/f_{+}$ is comparable with the ratio 
$n_{-}/|f_{-}|$. Otherwise, they are typically ``shallow'' and they do not 
represent sufficiently effective obstacles for the particle motion. The 
``trap-permeability'' parameter, as defined by 
$\theta\stackrel{{\rm def}}{=}n_{-}/|f_{-}|-n_{+}/f_{+}$, will play an 
important role in the following discussion \cite{monthus1,monthus2}. In fact, 
it is proportional to the mean force: $\theta=\mu(n_{-}+n_{+})/|f_{-}|f_{+}$. 
Having fixed the forces $f_{\pm}$, the mean force $\mu$ can be either positive 
or negative, depending on the parameters $n_{\pm}$, the value $\mu=0$ 
separating two regions with essentially different time-asymptotic dynamics. 

Let us consider again the small-$z$ limit of Eq.\ (\ref{ran24}). The 
quantities $q_{-}(z)$ and $q_{-}'(z)$ behave as in the preceding Subsection. 
Presently, however, one has $q_{+}(z)\sim z/f_{+}$ and 
$q_{+}'(z)\rightarrow-f_{+}$. Thereupon, the small-$z$ limit of the support 
(\ref{ran17}) is now the interval $[0,|f_{-}|]$. Further, the variable 
(\ref{ran27}) diverges and one has to use the analytic continuation of the 
hypergeometric function in Eq.\ (\ref{ran26}). 
%%%%%%%%%%%%%%%%%%%%%%%%%%%%%%%%%%%%%%%%%%%%%%%%%%%%%%%%%%%%%%%%%%%%%%%
\subsubsection{Negative mean force ($\mu<0$)}
%%%%%%%%%%%%%%%%%%%%%%%%%%%%%%%%%%%%%%%%%%%%%%%%%%%%%%%%%%%%%%%%%%%%%%%
In the small-$z$ limit the normalization constant $C(z)$ alone converges to a 
finite number and we can safely carry out this limit separately in $C(z)$ and 
in the rest of the expression (\ref{ran24}). The result reads
\begin{eqnarray}
\rule[-1ex]{0em}{6ex}\lim_{z\rightarrow0^{+}}\kappa(q;z)&=&
\frac{n_{+}}{n_{-}+n_{+}}
\frac{f_{+}^{\frac{n_{-}}{|f_{-}|}}
\left(|f_{-}|+f_{+}\right)^{\frac{n_{+}}{f_{+}}-\frac{n_{-}}{|f_{-}|}+1}}
{|f_{-}|^{\frac{n_{+}}{f_{+}}-1}}
\frac{1}{\displaystyle\sob\left(
\frac{n_{-}}{|f_{-}|},\frac{n_{+}}{f_{+}}-\frac{n_{-}}{|f_{-}|}\right)}
\times\nonumber\\
\label{dis2}
\rule[-1ex]{0em}{6ex}&\times&
q^{\frac{n_{+}}{f_{+}}-\frac{n_{-}}{|f_{-}|}-1}
\left(|f_{-}|-q\right)^{\frac{n_{-}}{|f_{-}|}-1}
\left(q+f_{+}\right)^{-\frac{n_{+}}{f_{+}}-1}
\Theta(q;0,|f_{-}|)\,.
\end{eqnarray}
All the moments of this limiting density exist and can be computed 
analytically. Let us just quote here the result for $k=1$: 
$\lim_{z\rightarrow0^{+}}\left\langle Q(z)\right\rangle=|\mu|$, {\em i.e.\/} 
we have again $\lim_{t\rightarrow\infty}\langle P(0,0;t)\rangle=|\mu|$. The 
mean trapping time diverges and the time-asymptotic velocity vanishes. As 
compared to the previous Subsection, the presence of traps does not modify the 
{\em modus\/} of the asymptotic dynamics. 
%%%%%%%%%%%%%%%%%%%%%%%%%%%%%%%%%%%%%%%%%%%%%%%%%%%%%%%%%%%%%%%%%%%%%%%
\subsubsection{Zero mean force ($\mu=0$)}
%%%%%%%%%%%%%%%%%%%%%%%%%%%%%%%%%%%%%%%%%%%%%%%%%%%%%%%%%%%%%%%%%%%%%%%
In this Sinai-like case, the small-$z$ analysis of the general expression for 
the first moment $\langle P(z)\rangle$ together with the Tauber theorem for 
the inverse Laplace transformation \cite{feller} yield the logarithmic decay
\begin{equation}
\label{dis6}
\left\langle P(0,0;t)\right\rangle\stackrel{t\rightarrow\infty}{\approx}
\frac{|f_{-}|f_{+}}{n_{-}+n_{+}}\,
\frac{1}{\log t}\,\,.
\end{equation}
%%%%%%%%%%%%%%%%%%%%%%%%%%%%%%%%%%%%%%%%%%%%%%%%%%%%%%%%%%%%%%%%%%%%%%%
\subsubsection{Positive mean force ($\mu>0$)}
%%%%%%%%%%%%%%%%%%%%%%%%%%%%%%%%%%%%%%%%%%%%%%%%%%%%%%%%%%%%%%%%%%%%%%%
In the small-$z$ limit, the normalization constant $C(z)$ diverges as 
$z^{-\theta}$, $\theta>0$. The limiting density (\ref{ran24}) is concentrated 
at one point: $\lim_{z\rightarrow0^{+}}\kappa(q;z)=\delta(q)$. All the moments 
$Q^{k}(z)$ are self-averaging quantities, their (non-random) limiting value 
being zero. On the other hand, the density $\lim_{z\rightarrow0^{+}}\pi(p;z)$ 
is a well behaved function, concentrated on the interval 
$[f_{+}^{-1},+\infty[$. Actually, if we first handle the substitution 
$\pi(p;z)=z\kappa(zp;z)$ in Eq.\ (\ref{ran24}) and {\em afterwards\/} exercise 
the small-$z$ limit, we get
\begin{eqnarray}
\rule[-1ex]{0em}{6ex}\lim_{z\rightarrow0^{+}}\pi(p;z)&=&
\frac{n_{-}}{n_{-}+n_{+}}\,
\frac{(|f_{-}|+f_{+})^{\frac{n_{-}}{|f_{-}|}-\frac{n_{+}}{f_{+}}+1}}
{|f_{-}|f_{+}^{\frac{n_{-}}{|f_{-}|}-\frac{n_{+}}{f_{+}}}}\,
\sob^{-1}\left(\frac{n_{+}}{f_{+}},\frac{n_{-}}{|f_{-}|}-
\frac{n_{+}}{f_{+}}\right)\times\nonumber\\
\label{dis3}
\rule[-1ex]{0em}{6ex}&\times&p
\left(p+\frac{1}{|f_{-}|}\right)^{-\frac{n_{-}}{|f_{-}|}-1}\,
\left(p-\frac{1}{f_{+}}\right)^{\frac{n_{+}}{f_{+}}-1}\,
\Theta\left(p;\frac{1}{f_{+}},+\infty\right)\,\,.
\end{eqnarray}
Here we come to an essential conclusion \cite{monthus2}: the moment 
$\lim_{z\rightarrow0^{+}}\langle P^{k}(z)\rangle$ is only finite if 
$k<\theta$. Specifically, for $\theta\in]0,1[$, the limit 
$\lim_{z\rightarrow0^{+}}\langle P(z)\rangle$ is infinite, {\em i.e.\/} the 
mean trapping time is also infinite and the time-asymptotic velocity is zero. 
More precisely, using again the Tauber theorem, we get
\begin{equation}
\label{dis4}
\left\langle P(0,0;t)\right\rangle\stackrel{t\rightarrow\infty}{\approx}
\frac{n_{-}}{n_{+}(n_{-}+n_{+})}\,
\frac{\Gamma^{2}\left(\displaystyle\frac{n_{-}}{|f_{-}|}\right)}
{\Gamma^{2}\left(\displaystyle\frac{n_{+}}{f_{+}}\right)\,\Gamma(\theta)}\,
\frac{f_{+}^{2(1-\theta)}(|f_{-}|+f_{+})^{2\theta}}{|f_{-}|^{2\theta}}\,
\frac{1}{t^{\theta}}\,\,.
\end{equation}
If $\theta\ge1$, the first moment 
$\lim_{z\rightarrow0^{+}}\left\langle P(z)\right\rangle$ is finite and its 
reciprocal gives the (self-averaging) time-asymptotic velocity:
\begin{equation}
\label{dis5}
\lim_{t\rightarrow\infty}\lim_{\lambda\rightarrow\infty}V(\lambda;t)=
(\theta-1)\frac{(n_{-}+n_{+})|f_{-}|f_{+}}
{(n_{-}+n_{+})^{2}-n_{-}|f_{-}|+n_{+}f_{+}}\,\,.
\end{equation}
However, if $\theta\in[1,2]$, the small-$z$ limit of the second moment 
$\lim_{z\rightarrow0^{+}}\left\langle P^{2}(z)\right\rangle$ is infinite. This 
can be shown to imply the vanishing of the (static) diffusion constant for the 
disorder averaged dynamics (this quantity is not discussed here). 
%%%%%%%%%%%%%%%%%%%%%%%%%%%%%%%%%%%%%%%%%%%%%%%%%%%%%%%%%%%%%%%%%%%%%%%
%%%%%%%%%%%%%%%%%%%%%%%%%%%%%%%%%%%%%%%%%%%%%%%%%%%%%%%%%%%%%%%%%%%%%%%
\subsection{Both forces are positive ($0<f_{-}<f_{+}$)}
%%%%%%%%%%%%%%%%%%%%%%%%%%%%%%%%%%%%%%%%%%%%%%%%%%%%%%%%%%%%%%%%%%%%%%%
%%%%%%%%%%%%%%%%%%%%%%%%%%%%%%%%%%%%%%%%%%%%%%%%%%%%%%%%%%%%%%%%%%%%%%%
In this case, the slope of the potential is always negative: the particle just 
slides towards infinity. The small-$z$ limit of the stationary probability 
density for the random variable $P(z)$ again follows from Eqs.\ (\ref{ran24}), 
(\ref{ran26}). The normalization constant alone tends to zero and one must 
operate with the whole expression (\ref{ran24}). The result reads
\begin{eqnarray}
\rule[-1ex]{0em}{6ex}\lim_{z\rightarrow0^{+}}\pi(p;z)&=&
\mu\left(\frac{f_{-}f_{+}}{f_{+}-f_{+}}\right)^
{\frac{n_{-}}{f_{-}}+\frac{n_{+}}{f_{+}}-1}
\sob^{-1}\left(\frac{n_{-}}{f_{-}},\frac{n_{+}}{f_{+}}\right)
\times\nonumber\\
\label{dis7}
\rule[-1ex]{0em}{6ex}&\times&p\,
\left(\frac{1}{f_{-}}-p\right)^{\frac{n_{-}}{f_{-}}-1}
\left(p-\frac{1}{f_{+}}\right)^{\frac{n_{+}}{f_{+}}-1}\,
\Theta\left(p;\frac{1}{f_{+}},\frac{1}{f_{-}}\right)\,\,.
\end{eqnarray}
Obviously enough, all the moments of this density exist. For $k=1$, we get
\begin{equation}
\label{dis8}
\tau=\lim_{z\rightarrow0^{+}}\langle P(z)\rangle=
\frac{n_{-}f_{-}+n_{+}f_{+}+(n_{-}+n_{+})^{2}}
{(n_{-}+n_{+})(n_{-}f_{+}+n_{+}f_{-}+f_{-}f_{+})}\,\,.
\end{equation}
As expected, the mean trapping time is finite. The time-asymptotic velocity is 
self-averaging and equals the reciprocal of the above expression
\cite{monthus1,monthus2}. There is no anomalous dynamical phase in this case. 
%%%%%%%%%%%%%%%%%%%%%%%%%%%%%%%%%%%%%%%%%%%%%%%%%%%%%%%%%%%%%%%%%%%%%%%
%%%%%%%%%%%%%%%%%%%%%%%%%%%%%%%%%%%%%%%%%%%%%%%%%%%%%%%%%%%%%%%%%%%%%%%
\subsection{White shot-noise limit}
%%%%%%%%%%%%%%%%%%%%%%%%%%%%%%%%%%%%%%%%%%%%%%%%%%%%%%%%%%%%%%%%%%%%%%%
%%%%%%%%%%%%%%%%%%%%%%%%%%%%%%%%%%%%%%%%%%%%%%%%%%%%%%%%%%%%%%%%%%%%%%%
Originally, the dichotomic quenched force has been described by four 
parameters, $n_{\pm}\ge 0$ and $f_{\pm}$. Another convenient equivalent 
four-parameter set is the mean force $\mu$, the intensity $\sigma$, the 
correlation length $\lambda_{c}=1/(n_{-}+n_{+})$, already introduced in 
Eq.\ (\ref{ran14}), and the ``non-Gaussianity'' parameter \cite{broeck1} 
$\gamma=|f_{+}-f_{-}|/(n_{-}+n_{+})$, the meaning of which being explained 
below. It is well known \cite{horsthemke,broeck1} that an appropriate limit of 
the dichotomic noise yields the Poisson white shot-noise. Actually, consider 
the parametrization
\begin{equation}
\label{dis9}
f_{-}=-\frac{\sigma-\gamma\mu}{\gamma}\,\,\,,\,\,\,
f_{+}=\xi\,\,\,,\,\,\,
n_{-}=\frac{\sigma}{\gamma^{2}}\,\,\,,\,\,\,
n_{+}=\frac{\xi}{\gamma}\,\,.
\end{equation}
If we increase the parameter $\xi$, the force $f_{+}$ increases and the mean 
length of the segments with the force $f_{+}$ tends to zero. In the limit 
$\xi\rightarrow\infty$, the parameters $\mu$, $\sigma$, and $\gamma$ keep 
their values, whereas the correlation length $\lambda_{c}$ tends to zero. The 
limiting form of the correlation function in Eq.\ (\ref{ran14}) is 
$2\sigma\delta(\lambda-\lambda')$. After the indicated limit, the quenched 
force displays an array of randomly positioned $\delta$-impulses on the 
constant background $-(\sigma-\gamma\mu)/\gamma$ ({\em c.f.\/} Fig.\ 3).
The mean (dimensionless) distance between the $\delta$-impulses is 
$\gamma^{2}/\sigma$, their weights being {\em randomly\/} distributed with the 
probability density $\gamma^{-1}\exp(-w/\gamma)\Theta(w)$. Thus the parameter 
$\gamma$ represents the mean weight of the impulses. On the whole, in the 
present Subsection, the random potential is described by the three parameters 
$\sigma\ge 0$, $\gamma\ge 0$, and $\mu$. The potential wells only exist if 
$f_{-}<0$, {\em i.e.\/} if $\gamma\mu<\sigma$; Fig.\ 3 illustrates the typical 
form of the potential in this case. 

Let us now carry out this limiting process in Eqs.\ (\ref{ran24}) and 
(\ref{ran26}). We get $q_{+}(z)\rightarrow 0$ and 
$\nu_{+}(z)\rightarrow1/\gamma$, {\em i.e.\/} the support of the density 
$\kappa(q;z)$ comes to be the interval $[0,q_{-}(z)]$. The density itself reads
\begin{eqnarray}
\rule[-1ex]{0em}{6ex}\kappa(q;z)&=&
\frac{q_{-}(z)^{\nu_{-}(z)+1/\gamma}}{[q_{-}(z)-q_{-}'(z)]^{\nu_{-}(z)+1}}
\sob^{-1}\left[\nu_{-}(z),1+1/\gamma\right]\times\nonumber\\
\rule[-1ex]{0em}{6ex}&\times&
\sof^{-1}\left[\nu_{-}(z),\nu_{-}(z)+1,\nu_{-}(z)+\frac{1}{\gamma};
\frac{q_{-}(z)}{q_{-}(z)-q_{-}'(z)}\right]
\times\nonumber\\
\label{dis10}
\rule[-1ex]{0em}{6ex}&\times&
q^{1/\gamma}[q_{-}(z)-q]^{\nu_{-}(z)-1}[q-q_{-}'(z)]^{-\nu_{-}(z)-1}
\Theta\left[q;0,q_{-}(z)\right]\,\,. 
\end{eqnarray}
We are again interested in the small-$z$ limit of this probability density and 
in its moments. We shall restrict the discussion to the physically interesting 
case with traps, {\em i.e.\/} $\gamma\mu<\sigma$. In this case, the variable 
of the Gauss hypergeometric function in Eq.\ (\ref{ran26}) tends to $1^{-}$ 
and we use an appropriate analytic-continuation formula \cite{abramowitz}.

If $\mu<0$, the time-asymptotic value of the averaged density at origin is 
again $\lim_{t\rightarrow\infty}\left\langle P(0,0;t)\right\rangle=|\mu|$. 
If $\mu=0$ is zero, we observe the logarithmic decay
\begin{equation}
\label{dis12}
\left\langle P(0,0;t)\right\rangle\stackrel{t\rightarrow\infty}{\approx}
\sigma\,\frac{1}{\log t}\,\,.
\end{equation}
If $\mu>0$, the time-asymptotics is controlled by the trap-permeability 
parameter $\theta$. Presently, it can be written as 
$\theta=\mu/(\sigma-\mu\gamma)$. For $\theta\in]0,1]$, the trapping time 
diverges and the asymptotic velocity vanishes. More precisely, we have 
\begin{equation}
\label{dis13}
\left\langle P(0,0;t)\right\rangle\stackrel{t\rightarrow\infty}{\approx}
\frac{\Gamma^{2}\left(\displaystyle\theta+\frac{1}{\gamma}\right)}
{\Gamma^{2}\left(\displaystyle\frac{1}{\gamma}\right)\Gamma(\theta)}\,
\frac{\sigma\gamma^{2\theta}}{(\sigma-\mu\gamma)^{2\theta}}\,
\frac{1}{t^{\theta}}\,\,.
\end{equation}
Finally, if $\theta\ge1$, the first moment 
$\lim_{z\rightarrow0^{+}}\left\langle P(z)\right\rangle$ is finite and its 
reciprocal gives the (self-averaging) time-asymptotic velocity:
\begin{equation}
\label{dis14}
\lim_{t\rightarrow\infty}\lim_{\lambda\rightarrow\infty}V(\lambda;t)=
(\theta-1)\frac{\sigma-\mu\gamma}{1+\gamma}\,\,.
\end{equation}
%%%%%%%%%%%%%%%%%%%%%%%%%%%%%%%%%%%%%%%%%%%%%%%%%%%%%%%%%%%%%%%%%%%%%%%
%%%%%%%%%%%%%%%%%%%%%%%%%%%%%%%%%%%%%%%%%%%%%%%%%%%%%%%%%%%%%%%%%%%%%%%
\subsection{Gaussian-white-noise limit}
%%%%%%%%%%%%%%%%%%%%%%%%%%%%%%%%%%%%%%%%%%%%%%%%%%%%%%%%%%%%%%%%%%%%%%%
%%%%%%%%%%%%%%%%%%%%%%%%%%%%%%%%%%%%%%%%%%%%%%%%%%%%%%%%%%%%%%%%%%%%%%%
It is well known \cite{horsthemke,broeck1} that an appropriate limit of the 
dichotomic noise yields the Gaussian white noise. However, the Gaussian white 
noise can be also obtained as a limit of the above-described white shot noise: 
one simply sets $\gamma\rightarrow0^{+}$. This means that the mean weight of 
the $\delta$-impulses of the force tends to zero, and simultaneously their 
density $n_{-}=\sigma/\gamma^{2}$ increases, such that the product 
(density)$\times$(mean weight)$^{2}$ remains constant. The bias $\mu$ and the 
intensity $\sigma$ keep their values and the correlation function in Eq.\ 
(\ref{ran14}) is again $2\sigma\delta(\lambda-\lambda')$. The quenched force 
displays an infinitely dense array of $\delta$ peaks in both directions, their 
weights being infinitely small. Notice that this limit can only be achieved if 
we start with the dichotomic random force taking two values of different 
signs. 

The Gaussian-white-noise results simply emerge after we carry out the 
small-$\gamma$ limit in Eqs.\ (\ref{dis10}). Particularly, we 
have $q_{-}(z)\rightarrow\infty$, {\em i.e.\/} the support of the probability 
density $\kappa(q;z)$ becomes the infinite interval $[0,+\infty[$. The density 
itself reads
\begin{equation}
\label{dis15}
\kappa(q;z)=\frac{1}{2}\frac{z^{\theta/2}}
{\displaystyle\sok_{\theta}\left(\frac{2}{\sigma}\sqrt{z}\right)}
\frac{1}{q^{\theta+1}}
\exp\left[-\frac{1}{\sigma}\left(q+\frac{z}{q}\right)\right]
\Theta(q;0,+\infty)\,\,,
\end{equation}
in accordance with the result found in Refs.\ \cite{bouchaud,aslangulg2}. 
Presently, the trap-permeability parameter simply measures the ratio between 
the mean force and its intensity: $\theta=\mu/\sigma$. 

The formula (\ref{dis15}) is valid for arbitrary values of the parameters 
$\sigma$ and $\mu$. If $\mu<0$, we have again 
$\lim_{t\rightarrow\infty}\left\langle P(0,0;t)\right\rangle=|\mu|$. In the 
Sinai case, {\em i.e.\/} for $\mu=0$, the asymptotic behaviour is again given 
by Eq.\ (\ref{dis12}). The corresponding result for the infinite line without 
the reflecting boundary condition at the origin is \cite{aslangulg2} 
$\left\langle P(0,0;t)\right\rangle\stackrel{t\rightarrow\infty}{\approx}
\sigma/(\log t)^{2}$. Thus the presence of the boundary slows down the decay 
of the disorder-averaged probability density at the origin. Having $\mu>0$ and 
$\theta\in]0,1[$, the asymptotic velocity vanishes and the averaged 
probability density at the origin decreases algebraically as 
\begin{equation}
\label{dis17}
\left\langle P(0,0;t)\right\rangle\stackrel{t\rightarrow\infty}{\approx}
\frac{\sigma^{1-2\theta}}{\Gamma(\theta)}\,\frac{1}{t^{\theta}}\,\,.
\end{equation}
For example, $\theta=1/2$ yields the exact solution 
$\left\langle P(0,0;t)\right\rangle=1/\sqrt{\pi t}$, valid for any time. 
Finally, when $\theta\ge1$, the time-asymptotic disorder-averaged mean 
position linearly increases, the self-averaging velocity being 
$(\theta-1)\sigma$. The damping of the disorder-averaged probability density 
at the origin can be exemplified by taking $\theta=3/2$: we then get 
$\langle P(z)\rangle=2/(\sigma+2\sqrt{z})$, that is
\begin{equation}
\label{dis18}
\langle P(0,0;t)\rangle=\frac{1}{\sqrt{\pi t}}-
\frac{\sigma}{2}\exp\left(\frac{1}{4}\sigma^{2}t\right)\,
{\rm erfc}\left(\frac{1}{2}\sigma\sqrt{t}\right)
\stackrel{t\rightarrow\infty}{\approx}
\frac{2}{\sigma^{2}\sqrt{\pi}}\,\frac{1}{t^{3/2}}\,\,.
\end{equation}
This asymptotic behaviour should be contrasted with the exponential damping 
which takes place in the presence of a positive homogeneous deterministic 
force, as found in Subsection 2.1.
%%%%%%%%%%%%%%%%%%%%%%%%%%%%%%%%%%%%%%%%%%%%%%%%%%%%%%%%%%%%%%%%%%%%%%%
%%%%%%%%%%%%%%%%%%%%%%%%%%%%%%%%%%%%%%%%%%%%%%%%%%%%%%%%%%%%%%%%%%%%%%%
%%%%%%%%%%%%%%%%%%%%%%%%%%%%%%%%%%%%%%%%%%%%%%%%%%%%%%%%%%%%%%%%%%%%%%%
\section{Conclusion}
%%%%%%%%%%%%%%%%%%%%%%%%%%%%%%%%%%%%%%%%%%%%%%%%%%%%%%%%%%%%%%%%%%%%%%%
%%%%%%%%%%%%%%%%%%%%%%%%%%%%%%%%%%%%%%%%%%%%%%%%%%%%%%%%%%%%%%%%%%%%%%%
%%%%%%%%%%%%%%%%%%%%%%%%%%%%%%%%%%%%%%%%%%%%%%%%%%%%%%%%%%%%%%%%%%%%%%%
In the present paper, a transfer-matrix-like method for solving diffusion 
problems in a piecewise linear random potential has been introduced. The 
formulae for the Green function derived in the second Section can be easily 
adapted to numerical simulation of the diffusive motion in {\em any\/} 
potential of the mentioned type. For example, the force can be assumed to be a 
semi-Markovian or a non-Markovian variant of the dichotomic noise 
\cite{araujo}-\cite{west}, it can exhibit jumps of random magnitudes (kangaroo 
process \cite{feller}), etc. For any such process, our analysis is valid up to 
Subsection 3.3. Our subsequent choice of a Markovian dichotomic process has 
been dictated by a relatively direct possibility to get the asymptotic 
solution of the stochastic equations (\ref{ran5}), (\ref{ran6}). 

Let us summarize the preceding discussion. The dynamical effects of the 
quenched disorder have been evinced by examining the varying stochastic 
features of a single random variable, namely, the probability density of the 
particle's occurence at the origin. Having a negative mean bias, the 
time-asymptotic and disorder-averaged value of this quantity is proportional 
to the absolute value of the mean force. In the Sinai-like case, {\em i.e.\/} 
for the vanishing mean bias, we have given the exact asymptotic formula 
describing the decay of this quantity. The decay is slower than in the 
corresponding model without the reflecting boundary at the origin. Finally, 
having a positive mean bias, the particle escapes towards infinity, with a 
finite velocity or not, depending on the value of the trap-permeability 
parameter. The existence of deep traps with a long trapping time is crucial 
for the existence of anomalous dynamical phases.

In the present work, we have chosen to formulate the diffusion problem in the 
presence of reflecting boundary conditions. After a slight modification, the 
method can be adapted to other types of boundary conditions. For instance, 
taking a fixed probability density at two boundaries, our method can be used 
to the analysis of the stationary-flux distribution in the one-dimensional 
random medium \cite{monthus1}. 

We have not aimed at the exhaustive description of the particle dynamics. 
Instead, we have concentrated on features which can be directly related to the 
probability density at the origin. The detailed description of the ``noise''
$P(\lambda;z)$ allows, at least in principle, for an investigation of other 
aspects, such as the time-asymptotic thermally-averaged first moment of the 
particle's position. Another example would be the disorder-averaged diffusion 
coefficient. Its analysis requires the calculation of the second 
thermally-averaged moment of the particle's position on the one hand, and the 
higher-order (generally non-self-averaging) terms in the small-$z$ expansion 
of the first moment on the other hand. We have shown that the 
thermally-averaged moments obey a system of stochastic differential equations 
with $P(\lambda;z)$ playing the role of the ``input'' noise. The probabilistic 
description of the moments will be reported elsewhere. 

Summing up, the paper presents an approximation-free study of the diffusive 
dynamics in an one-dimensional Markovian Poisson random potential. It provides 
a firm basis for the intuitive understanding of diffusion in more involved 
circumstances.
%%%%%%%%%%%%%%%%%%%%%%%%%%%%%%%%%%%%%%%%%%%%%%%%%%%%%%%%%%%%%%%%%%%%%%%
\subsection*{Acknowledgments.}
One of the authors (P.\ Ch.) would like to thank the University Paris VII for
financial support and to express his gratitude for the hospitality extended to 
him at the ``Groupe de Physique des Solides''.
%%%%%%%%%%%%%%%%%%%%%%%%%%%%%%%%%%%%%%%%%%%%%%%%%%%%%%%%%%%%%%%%%%%%%%%
%%%%%%%%%%%%%%%%%%%%%%%%%%%%%%%%%%%%%%%%%%%%%%%%%%%%%%%%%%%%%%%%%%%%%%%
%%%%%%%%%%%%%%%%%%%%%%%%%%%%%%%%%%%%%%%%%%%%%%%%%%%%%%%%%%%%%%%%%%%%%%%
\appendix
\section*{Appendix}
\setcounter{equation}{0}
\renewcommand{\theequation}{A.\arabic{equation}}
Let us consider the integrals
\begin{eqnarray}
\rule[-1ex]{0em}{6ex}I_{k}(z)&\stackrel{{\rm def}}{=}&
\int_{q_{+}(z)}^{q_{-}(z)}dq\,q^{k}\,\left\{
\frac{1}{[q_{-}(z)-q][q-q_{-}'(z)]}+\frac{1}{[q-q_{+}(z)][q-q_{+}'(z)]}
\right\}\times\nonumber\\
\label{app1}
\rule[-1ex]{0em}{6ex}&\times&
\left[\frac{q_{-}(z)-q}{q-q_{-}'(z)}\right]^{\nu_{-}(z)}
\left[\frac{q-q_{+}(z)}{q-q_{+}'(z)}\right]^{\nu_{+}(z)}\,\,.
\end{eqnarray}
Hence $I_{0}(z)$ is equal to the integration constant $C(z)$ in Eq.\ 
(\ref{ran17}). Having known $I_{0}(z)$ and $I_{k}(z)$ for some $k\ge1$, 
the $k$-th stationary moment $\langle Q^{k}(z)\rangle$ is simply given by the 
ratio $I_{k}(z)/I_{0}(z)$. In this Appendix, we want to display some 
intermediate steps in the calculation of the above integrals. In order to keep 
the following formulae in a reasonable shape, we shall write $\nu_{\pm}$ 
instead of the more descriptive designation $\nu_{\pm}(z)$, used in the main 
text. Similar remarks hold for the four $z$-dependent quantities $q_{\pm}$ 
and $q_{\pm}'$.

First, we notice the equality 
\begin{equation}
\label{app2}
q=\frac{(q_{-}-q)(q-q_{-}')+(q-q_{+})(q-q_{+}')}{f_{+}-f_{-}}\,\,,
\end{equation}
and we introduce the substitution $x=(q-q_{+})/(q_{-}-q_{+})$. Thereupon, the 
integrals $I_{k}(z)$ assume the form
\begin{eqnarray}
\rule[-1ex]{0em}{6ex}I_{k}(z)&=&
\frac{(q_{-}-q_{+})^{2k-1}}{(f_{+}-f_{-})^{k}}
\int_{0}^{1}dx\,\left[(1-x)(x+a)+x(x+b)\right]^{k+1}\times\nonumber\\
\label{app3}
\rule[-1ex]{0em}{6ex}&\times&
\frac{(1-x)^{\nu_{-}-1}}{(x+a)^{\nu_{-}+1}}
\frac{x^{\nu_{+}-1}}{(x+b)^{\nu_{+}+1}}\,\,,
\end{eqnarray}
where $a=(q_{+}-q_{-}')/(q_{-}-q_{+})$ and $b=(q_{+}-q_{+}')/(q_{-}-q_{+})$.

Then, using the binomial theorem, we expand the $(k+1)$-th power 
in (\ref{app3}) and write $I_{k}(z)$ as a sum of $k+2$ integrals:
\begin{eqnarray}
\label{app4}
\rule[-1ex]{0em}{6ex}I_{k}(z)&=&
\frac{(q_{-}-q_{+})^{2k-1}}{(f_{+}-f_{-})^{k}}
\sum_{j=0}^{k+1}\frac{(k+1)!}{j!(k+1-j)!}J_{k,j}(z)\,\,,\\
\label{app5}
\rule[-1ex]{0em}{6ex}J_{k,j}(z)&=&\int_{0}^{1}dx\,
(1-x)^{\nu_{-}+j-1}(x+a)^{-\nu_{-}+j-1}x^{\nu_{+}+k-j}(x+b)^{-\nu_{+}+k-j}.
\,\,\,\,\,\,
\end{eqnarray}
Finally, the integrals $J_{k,j}(z)$ are expressed through the Appell function 
$\sof_{1}$ \cite{appell} (the hypergeometric function of two variables 
\cite{abramowitz,gradshteyn}). We have
\begin{eqnarray}
\rule[-1ex]{0em}{6ex}
J_{k,j}(z)&=&a^{-\nu_{-}+j-1}b^{-\nu_{+}+k-j}
\frac{\Gamma(\nu_{-}+j)\Gamma(\nu_{+}+k-j+1)}{\Gamma(\nu+k+1)}\times
\nonumber\\
\label{app6}
\rule[-1ex]{0em}{6ex}&\times&
\sof_{1}\left(\alpha,\beta,\beta',\gamma;-a^{-1},-b^{-1}\right)\,\,,
\end{eqnarray}
with $\alpha=\nu_{+}+k-j+1$, $\beta=\nu_{-}-j+1$, $\beta'=\nu_{+}-k+j$, 
$\gamma=\nu+k+1$, and $\nu=\nu_{-}+\nu_{+}$. In the small-$z$ limit, the 
behaviour of the result depends on the two variables $a$ and $b$; one can 
employ the functional relations between the Appell functions. Moreover, having 
a special relationship between the four parameters, the Appell function can be 
expressed in terms of the ordinary hypergeometric function. In our context, 
the case $k=0$ is exceptional due to the possibility of such reduction. 
Alternatively, this point can be seen by introducing the second substitution, 
$y=[x(1+b)]/(x+b)$. One then obtains
\begin{eqnarray}
\rule[-1ex]{0em}{6ex}J_{0,0}(z)&=&
\frac{b}{a^{\nu_{-}+1}(1+b)^{\nu_{+}+1}}
\int_{0}^{1}dy\,y^{\nu_{+}}(1-y)^{\nu_{-}-1}(1+uy)^{-\nu_{-}-1}=\nonumber\\
\label{app7}
\rule[-1ex]{0em}{6ex}&=&
\frac{(1+a)^{-1}}{a^{\nu_{-}}(1+b)^{\nu_{+}}}
\frac{\Gamma(\nu_{-})\Gamma(\nu_{+}+1)}{\Gamma(\nu+1)}\,
\sof(\nu_{-},\nu_{+},\nu+1;-u)\,\,,\\
\rule[-1ex]{0em}{6ex}J_{0,1}(z)&=&
\frac{1}{ba^{\nu_{-}}(1+b)^{\nu_{+}}}
\int_{0}^{1}dy\,y^{\nu_{+}-1}(1-y)^{\nu_{-}}(1+uy)^{-\nu_{-}}=\nonumber\\
\label{app8}
\rule[-1ex]{0em}{6ex}&=&
\frac{b^{-1}}{a^{\nu_{-}}(1+b)^{\nu_{+}}}
\frac{\Gamma(\nu_{-}+1)\Gamma(\nu_{+})}{\Gamma(\nu+1)}\,
\sof(\nu_{-},\nu_{+},\nu+1;-u)\,\,,
\end{eqnarray}
with $\nu=\nu_{-}+\nu_{+}$ and $u=(b-a)/[a(1+b)]$. Summing these two 
expressions, we acquire
\begin{equation}
\label{app9}
I_{0}(z)=\frac{J_{0,0}(z)+J_{0,1}(z)}{q_{-}-q_{+}}=\frac{1}{\Omega}
\,\frac{1}{a^{\nu_{-}}(1+b)^{\nu_{+}}}\,
\sob(\nu_{-},\nu_{+})\,\sof(\nu_{-},\nu_{+},\nu+1;-u)\,, 
\end{equation}
where $\sob(x,y)$ is the Euler beta function \cite{abramowitz} and the factor 
\begin{equation}
\label{app10}
\Omega=\frac{n_{+}(q_{-}-q_{-}')+n_{-}(q_{+}-q_{+}')}{n_{-}+n_{+}}=
\frac{n_{+}\sqrt{4z+f_{-}^{2}}+n_{-}\sqrt{4z+f_{+}^{2}}}{n_{-}+n_{+}}
\end{equation}
can be regarded as the mean value 
$\left\langle\sqrt{4z+\phi^{2}(\lambda)}\right\rangle$---{\em c.f.\/} 
Eq.\ (\ref{ran1}). Finally, upon inserting the definitions of the 
$z$-dependent quantities $a=a(z)$, $b=b(z)$ and $u=u(z)$, one recovers the 
normalization constant (\ref{ran19}) from the main text. 
%%%%%%%%%%%%%%%%%%%%%%%%%%%%%%%%%%%%%%%%%%%%%%%%%%%%%%%%%%%%%%%%%%%%%
%%%%%%%%%%%%%                                                    %%%%
%%%%%%%%%%%%%                  References.                       %%%%
%%%%%%%%%%%%%                                                    %%%%
%%%%%%%%%%%%%%%%%%%%%%%%%%%%%%%%%%%%%%%%%%%%%%%%%%%%%%%%%%%%%%%%%%%%%

%%%%%%%%%%%%%%%%%%%%%%%%%%%%%%%%%%%%%%%%%%%%%%%%%%%%%%%%%%%%%%%%%%%%%
%%%%%%%%%%%%%%%%%%%%%%%%%%%%%%%%%%%%%%%%%%%%%%%%%%%%%%%%%%%%%%%%%%%%%
%%%%%%%%%%%%%%%%%%%%%%%%%%%%%%%%%%%%%%%%%%%%%%%%%%%%%%%%%%%%%%%%%%%%%
%%%%%%%%%%%%%                                                    %%%%
%%%%%%%%%%%%%                  Figures.                          %%%%
%%%%%%%%%%%%%                                                    %%%%
%%%%%%%%%%%%%%%%%%%%%%%%%%%%%%%%%%%%%%%%%%%%%%%%%%%%%%%%%%%%%%%%%%%%%
\newpage
\pagestyle{empty}
\subsection*{Figure 1:}
\begin{figure}[h]
\begin{center}
\unitlength=1.00mm
\begin{picture}(140.00,140.00)
%%%%%%%%%%%%%%%%%%%%%%%%%%%%%%%%%%%%%%
\linethickness{0.1pt}
\thinlines
\put(0.0,0.0){\line(1,0){140.0}}
\put(140.0,0.0){\line(0,1){140.0}}
\put(140.0,140.0){\line(-1,0){140.0}}
\put(0.0,140.0){\line(0,-1){140.0}}
%%%%%%%%%%%%%%%%%%%%%%%%%%%%%%%%%%%%%%
\linethickness{0.5pt}
\thicklines
\put(35,35){\line(1,0){84}}
\put(120,35){\vector(1,0){0}}
\put(96.5,30){\makebox(27.5,0)[bl]{\rm Coordinate $x$}}
\put(35,91){\line(1,0){84}}
\put(120,91){\vector(1,0){0}}
\put(96.5,86){\makebox(27.5,0)[bl]{\rm Coordinate $x$}}
\put(35,14){\line(0,1){49}}
\put(35,64){\vector(0,1){0}}
\put(35,65){\makebox(0,0){\rm Potential $U(x)$}}
\put(35,73.5){\line(0,1){49}}
\put(35,123.5){\vector(0,1){0}}
\put(35,124.5){\makebox(0,0){\rm Force $f(x)$}}
%%%%%%%%%%%%%%%%%%%%%%%%%%%%%%%%%%%%%%
\thinlines
\multiput(35,77)(2,0){43}{\line(1,0){1}}
\put(33,77){\makebox(0,0){$f_{-}$}}
\put(33,91){\makebox(0,0){$0$}}
\multiput(35,119)(2,0){43}{\line(1,0){1}}
\put(33,119){\makebox(0,0){$f_{+}$}}
\multiput(35,15)(0,1){48}{\line(-1,-2){2}}
%%%%%%%%%%%%%%%%%%%%%%%%%%%%%%%%%%%%%%
\put(49.3,92){\makebox(1.6,0)[bl]{$x_{1}$}}
\put(77.3,92){\makebox(1.6,0)[bl]{$x_{2}$}}
\put(84.3,92){\makebox(1.6,0)[bl]{$x_{3}$}}
\put(105.3,92){\makebox(1.6,0)[bl]{$x_{4}$}}
\put(41,94){\makebox(1.57,0)[bl]{$l_{1}$}}
\put(62,94){\makebox(1.57,0)[bl]{$l_{2}$}}
\put(93.5,94){\makebox(1.57,0)[bl]{$l_{4}$}}
\put(111,94){\makebox(1.57,0)[bl]{$l_{5}$}}
%%%%%%%%%%%%%%%%%%%%%%%%%%%%%%%%%%%%%%
\multiput(49,14)(0,2){53}{\line(0,1){1}}
\multiput(77,14)(0,2){53}{\line(0,1){1}}
\multiput(84,14)(0,2){53}{\line(0,1){1}}
\multiput(105,14)(0,2){53}{\line(0,1){1}}
\thicklines
\put(49,90.5){\line(0,1){1}}
\put(77,90.5){\line(0,1){1}}
\put(84,90.5){\line(0,1){1}}
\put(105,90.5){\line(0,1){1}}
%%%%%%%%%%%%%%%%%%%%%%%%%%%%%%%%%%%%%%
\put(35,77){\line(1,0){14}}
\put(77,77){\line(1,0){7}}
\put(105,77){\line(1,0){15}}
\put(49,119){\line(1,0){28}}
\put(84,119){\line(1,0){21}}
%%%%%%%%%%%%%%%%%%%%%%%%%%%%%%%%%%%%%%
\put(35,55){\line(2,1){14}}
\put(49,62){\line(1,-1){28}}
\put(77,34){\line(2,1){7}}
\put(84,37.5){\line(1,-1){21}}
\put(105,16.5){\line(2,1){14}}
%%%%%%%%%%%%%%%%%%%%%%%%%%%%%%%%%%%%%%
\end{picture}
\end{center}
\caption{\label{fig1}
Typical realization of the Markovian Poisson dichotomic force taking two 
values of opposite signs, $f_{-}<0<f_{+}$ with $f_{+}=2|f_{-}|$. The slope of 
the corresponding potential is alternately positive and negative: there are 
traps in this case.}
\end{figure}
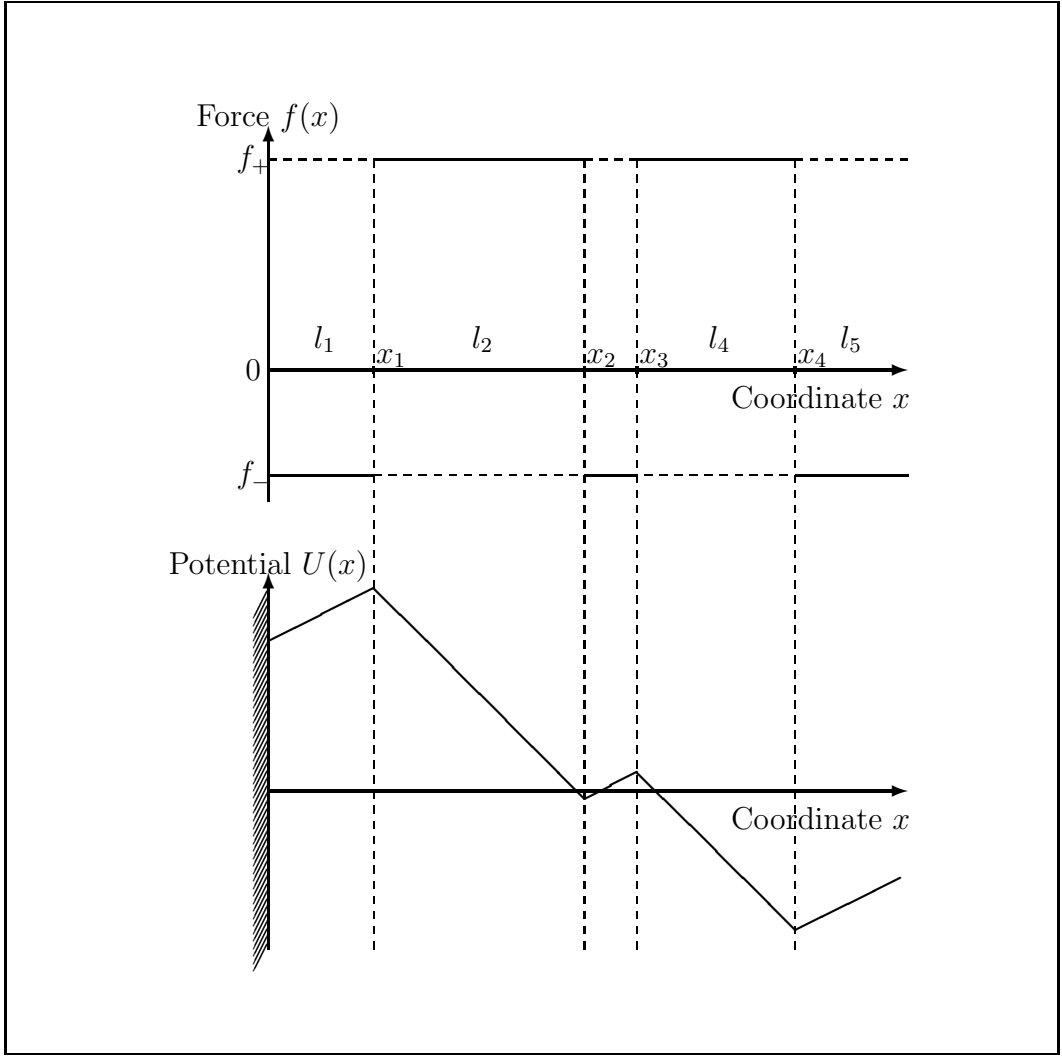
%%%%%%%%%%%%%%%%%%%%%%%%%%%%%%%%%%%%%%%%%%%%%%%%%%%%%%%%%%%%%%%%%%%%%
%%%%%%%%%%%%%%%%%%%%%%%%%%%%%%%%%%%%%%%%%%%%%%%%%%%%%%%%%%%%%%%%%%%%%
%%%%%%%%%%%%%%%%%%%%%%%%%%%%%%%%%%%%%%%%%%%%%%%%%%%%%%%%%%%%%%%%%%%%%
\newpage
\pagestyle{empty}
\subsection*{Figure 2:}
\begin{figure}[h]
\begin{center}
\unitlength=1.00mm
\begin{picture}(140.00,140.00)
%%%%%%%%%%%%%%%%%%%%%%%%%%%%%%%%%%%%%%
\linethickness{0.1pt}
\thinlines
\put(0.0,0.0){\line(1,0){140.0}}
\put(140.0,0.0){\line(0,1){140.0}}
\put(140.0,140.0){\line(-1,0){140.0}}
\put(0.0,140.0){\line(0,-1){140.0}}
%%%%%%%%%%%%%%%%%%%%%%%%%%%%%%%%%%%%%%
\linethickness{0.5pt}
\thicklines
\put(35,35){\line(1,0){84}}
\put(120,35){\vector(1,0){0}}
\put(96.5,30){\makebox(27.5,0)[bl]{\rm Coordinate $x$}}
\put(35,91){\line(1,0){84}}
\put(120,91){\vector(1,0){0}}
\put(96.5,86){\makebox(27.5,0)[bl]{\rm Coordinate $x$}}
\put(35,14){\line(0,1){49}}
\put(35,64){\vector(0,1){0}}
\put(35,65){\makebox(0,0){\rm Potential $U(x)$}}
\put(35,73.5){\line(0,1){49}}
\put(35,123.5){\vector(0,1){0}}
\put(35,124.5){\makebox(0,0){\rm Force $f(x)$}}
%%%%%%%%%%%%%%%%%%%%%%%%%%%%%%%%%%%%%%
\thinlines
\multiput(35,77)(2,0){43}{\line(1,0){1}}
\put(33,77){\makebox(0,0){$f_{-}$}}
\put(33,91){\makebox(0,0){$0$}}
\multiput(35,84)(2,0){43}{\line(1,0){1}}
\put(33,84){\makebox(0,0){$f_{+}$}}
\multiput(35,15)(0,1){48}{\line(-1,-2){2}}
%%%%%%%%%%%%%%%%%%%%%%%%%%%%%%%%%%%%%%
\put(47,92){\makebox(1.6,0)[bl]{$x_{1}$}}
\put(75,92){\makebox(1.6,0)[bl]{$x_{2}$}}
\put(82,92){\makebox(1.6,0)[bl]{$x_{3}$}}
\put(103,92){\makebox(1.6,0)[bl]{$x_{4}$}}
\put(41,94){\makebox(1.57,0)[bl]{$l_{1}$}}
\put(62,94){\makebox(1.57,0)[bl]{$l_{2}$}}
\put(93.5,94){\makebox(1.57,0)[bl]{$l_{4}$}}
\put(111,94){\makebox(1.57,0)[bl]{$l_{5}$}}
%%%%%%%%%%%%%%%%%%%%%%%%%%%%%%%%%%%%%%
\multiput(49,14)(0,2){39}{\line(0,1){1}}
\multiput(77,14)(0,2){39}{\line(0,1){1}}
\multiput(84,14)(0,2){39}{\line(0,1){1}}
\multiput(105,14)(0,2){39}{\line(0,1){1}}
%%%%%%%%%%%%%%%%%%%%%%%%%%%%%%%%%%%%%%
\thicklines
\put(49,90.5){\line(0,1){1}}
\put(77,90.5){\line(0,1){1}}
\put(84,90.5){\line(0,1){1}}
\put(105,90.5){\line(0,1){1}}
%%%%%%%%%%%%%%%%%%%%%%%%%%%%%%%%%%%%%%
\put(35,77){\line(1,0){14}}
\put(77,77){\line(1,0){7}}
\put(105,77){\line(1,0){15}}
\put(49,84){\line(1,0){28}}
\put(84,84){\line(1,0){21}}
%%%%%%%%%%%%%%%%%%%%%%%%%%%%%%%%%%%%%%
\put(35,22){\line(2,1){14}}
\put(49,29){\line(4,1){28}}
\put(77,36){\line(2,1){7}}
\put(84,39.5){\line(4,1){21}}
\put(105,44.75){\line(2,1){14}}
%%%%%%%%%%%%%%%%%%%%%%%%%%%%%%%%%%%%%%
\end{picture}
\end{center}
\caption{\label{fig2}
Typical realization of the Markovian Poisson dichotomic force taking two 
values of the negative signs, $f_{-}<f_{+}<0$ with $|f_{-}|=2|f_{+}|$. The 
slope of the corresponding potential is always positive. There are no traps in 
this case, the particle being stuck towards the reflecting boundary at the 
origin.}
\end{figure}
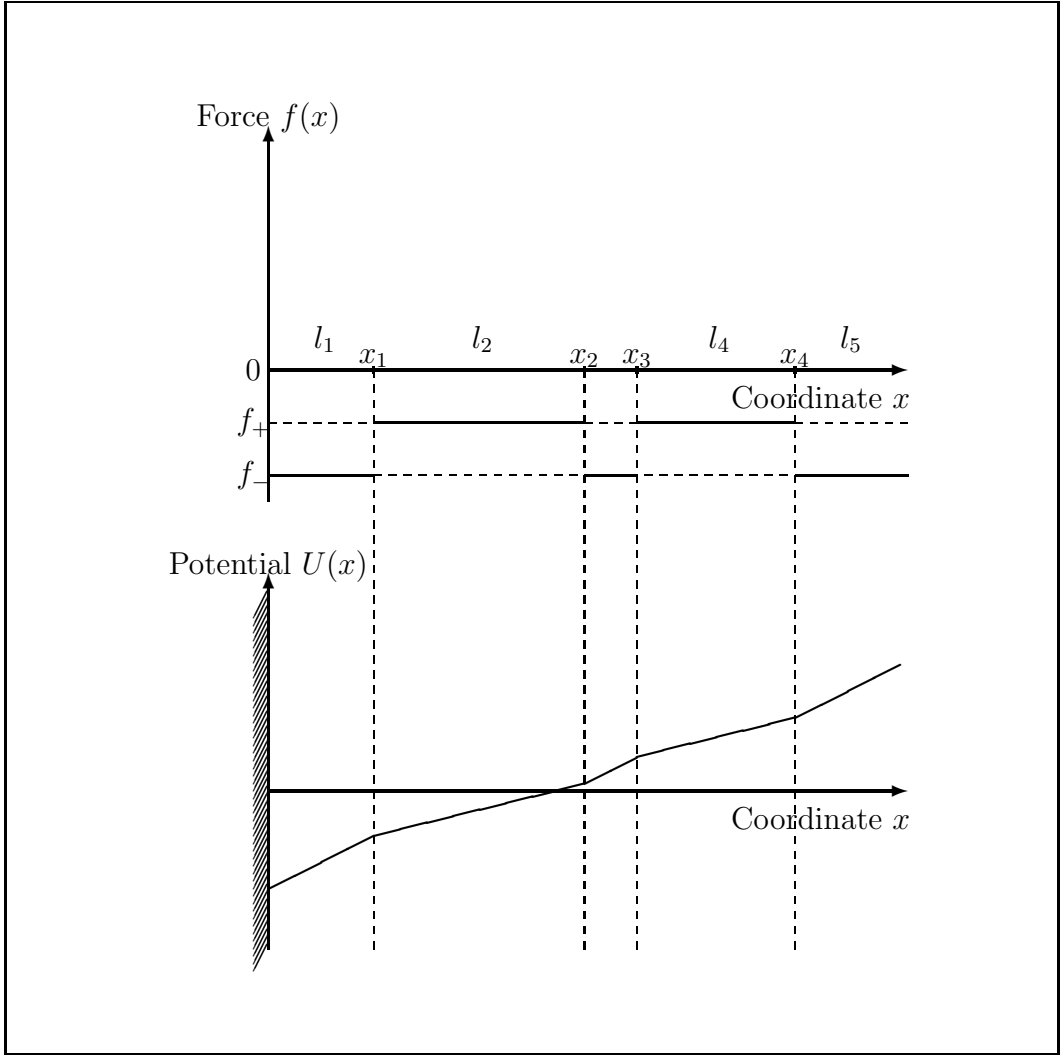
%%%%%%%%%%%%%%%%%%%%%%%%%%%%%%%%%%%%%%%%%%%%%%%%%%%%%%%%%%%%%%%%%%%%%
%%%%%%%%%%%%%%%%%%%%%%%%%%%%%%%%%%%%%%%%%%%%%%%%%%%%%%%%%%%%%%%%%%%%%
%%%%%%%%%%%%%%%%%%%%%%%%%%%%%%%%%%%%%%%%%%%%%%%%%%%%%%%%%%%%%%%%%%%%%
\newpage
\pagestyle{empty}
\subsection*{Figure 3:}
\begin{figure}[h]
\begin{center}
\unitlength=1.00mm
\begin{picture}(140.00,140.00)
%%%%%%%%%%%%%%%%%%%%%%%%%%%%%%%%%%%%%%
\linethickness{0.1pt}
\thinlines
\put(0.0,0.0){\line(1,0){140.0}}
\put(140.0,0.0){\line(0,1){140.0}}
\put(140.0,140.0){\line(-1,0){140.0}}
\put(0.0,140.0){\line(0,-1){140.0}}
%%%%%%%%%%%%%%%%%%%%%%%%%%%%%%%%%%%%%%
\linethickness{0.5pt}
\thicklines
\put(35,35){\line(1,0){84}}
\put(120,35){\vector(1,0){0}}
\put(106.5,30){\makebox(27.5,0)[bl]{\rm Coordinate $x$}}
\put(35,91){\line(1,0){84}}
\put(120,91){\vector(1,0){0}}
\put(96.5,86){\makebox(27.5,0)[bl]{\rm Coordinate $x$}}
\put(35,14){\line(0,1){49}}
\put(35,64){\vector(0,1){0}}
\put(35,65){\makebox(0,0){\rm Potential $U(x)$}}
\put(35,73.5){\line(0,1){49}}
\put(35,123.5){\vector(0,1){0}}
\put(35,124.5){\makebox(0,0){\rm Force $f(x)$}}
%%%%%%%%%%%%%%%%%%%%%%%%%%%%%%%%%%%%%%
\thinlines
\put(33,77){\makebox(0,0){$f_{-}$}}
\put(33,91){\makebox(0,0){$0$}}
\multiput(35,15)(0,1){48}{\line(-1,-2){2}}
%%%%%%%%%%%%%%%%%%%%%%%%%%%%%%%%%%%%%%
\put(49.3,92){\makebox(1.6,0)[bl]{$x_{1,2}$}}
\put(70,92){\makebox(1.6,0)[bl]{$x_{3,4}$}}
\put(84.3,92){\makebox(1.6,0)[bl]{$x_{5,6}$}}
\put(105.3,92){\makebox(1.6,0)[bl]{$x_{7,8}$}}
\put(41,94){\makebox(1.57,0)[bl]{$l_{1}$}}
\put(62,94){\makebox(1.57,0)[bl]{$l_{3}$}}
\put(79.5,94){\makebox(1.57,0)[bl]{$l_{5}$}}
\put(93.5,94){\makebox(1.57,0)[bl]{$l_{7}$}}
\put(114,94){\makebox(1.57,0)[bl]{$l_{9}$}}
%%%%%%%%%%%%%%%%%%%%%%%%%%%%%%%%%%%%%%
\multiput(49,14)(0,2){39}{\line(0,1){1}}
\multiput(77,14)(0,2){39}{\line(0,1){1}}
\multiput(84,14)(0,2){39}{\line(0,1){1}}
\multiput(105,14)(0,2){39}{\line(0,1){1}}
%%%%%%%%%%%%%%%%%%%%%%%%%%%%%%%%%%%%%%
\thicklines
\put(35,77){\line(1,0){85}}
\put(49,91){\line(0,1){7}}
\put(49,99){\vector(0,1){0}}
\put(77,91){\line(0,1){28}}
\put(77,120){\vector(0,1){0}}
\put(84,91){\line(0,1){14}}
\put(84,106){\vector(0,1){0}}
\put(105,91){\line(0,1){21}}
\put(105,113){\vector(0,1){0}}
%%%%%%%%%%%%%%%%%%%%%%%%%%%%%%%%%%%%%%
\put(35,52){\line(2,1){14}}
\put(49,59){\line(0,-1){7}}
\put(49,52){\line(2,1){28}}
\put(77,66){\line(0,-1){28}}
\put(77,38){\line(2,1){7}}
\put(84,41.5){\line(0,-1){14}}
\put(84,27.5){\line(2,1){21}}
\put(105,38){\line(0,-1){21}}
\put(105,17){\line(2,1){15}}
%%%%%%%%%%%%%%%%%%%%%%%%%%%%%%%%%%%%%%
\end{picture}
\end{center}
\caption{\label{fig3}
Typical realization of the white shot-noise random force and of the 
corresponding potential. The lengths of the vertical segments of the potential 
are independent, exponentially distributed random variables, their mean value 
being $\gamma$. There are traps in this case.}
\end{figure}
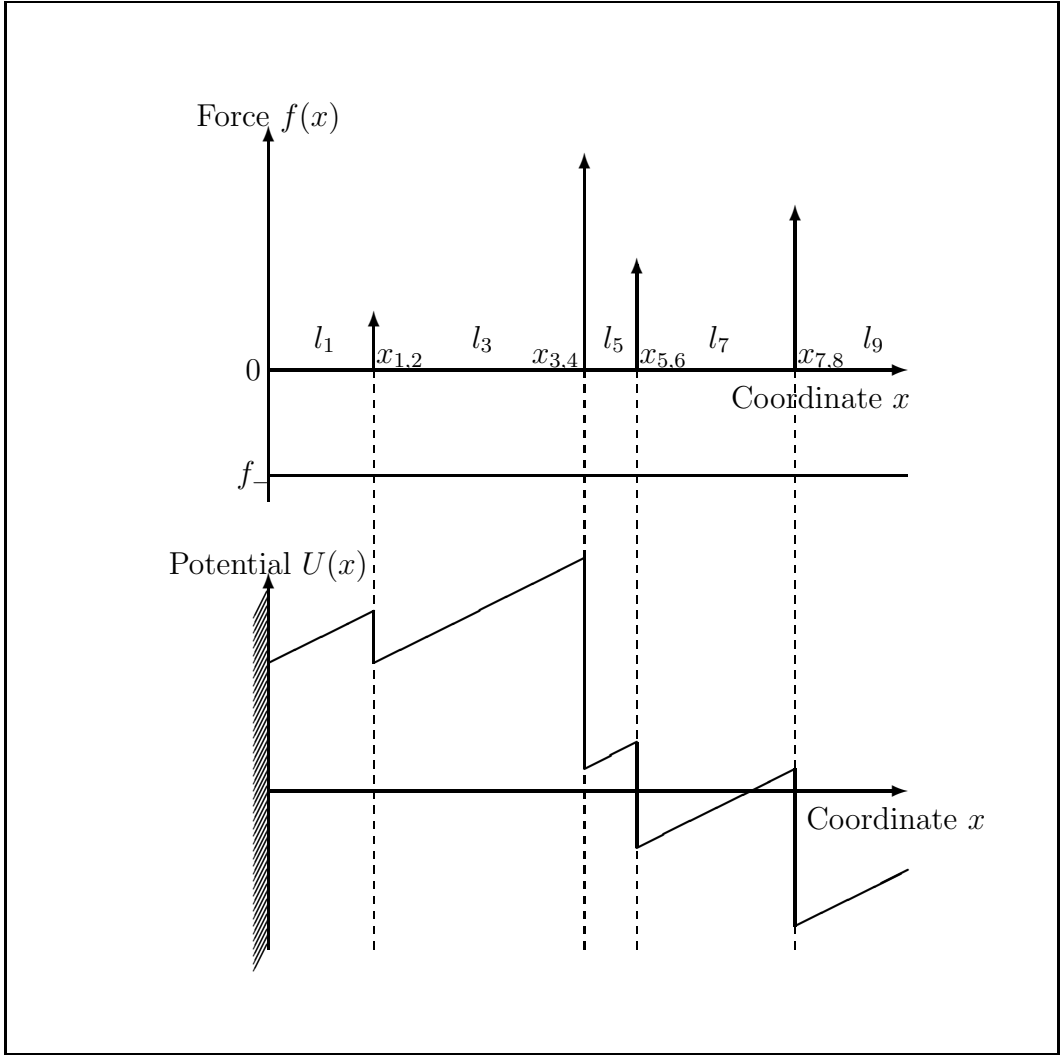
%%%%%%%%%%%%%%%%%%%%%%%%%%%%%%%%%%%%%%%%%%%%%%%%%%%%%%%%%%%%%%%%%%%%%
\end{document}